\documentclass[11pt,a4paper]{article}
\pdfoutput=1
\usepackage{jheppub}
\usepackage{epsfig}
\usepackage{latexsym}
\usepackage{amsfonts}
\usepackage{amsmath}
\usepackage{amsthm}
\usepackage{amssymb}
\usepackage{amsbsy}
\usepackage{multirow}
\usepackage{slashed}
\usepackage{extarrows}

\newcommand{\RR}{\mathbb{R}} 

\newcommand{\G}{\mathcal{G}}

\def\cala         {{\cal A}}

\def\cale         {{\cal E}}

\def\calk         {{\cal K}}
\def\call         {{\cal L}}
\def\calm         {{\cal M}}

\def\calo         {{\cal O}}
\def\calp         {{\cal P}}

\def\calr         {{\cal R}}

\def\calw         {{\cal W}}
\newsavebox{\uuunit}
\sbox{\uuunit}
    {\setlength{\unitlength}{0.825em}
     \begin{picture}(0.6,0.7)
        \thinlines
        \put(0,0){\line(1,0){0.5}}
        \put(0.15,0){\line(0,1){0.7}}
        \put(0.35,0){\line(0,1){0.8}}
       \multiput(0.3,0.8)(-0.04,-0.02){12}{\rule{0.5pt}{0.5pt}}
     \end {picture}}

\def\be{\begin{equation}}
\def\ee{\end{equation}}
\def\bea{\begin{eqnarray}}
\def\eea{\end{eqnarray}}

\def\a{\alpha}
\def\b{\beta}

\def\g{\gamma}
\def\G{\Gamma}
\def\d{\delta}
\def\e{\epsilon}
\def\D{\Delta}
\def\l{\lambda}
\def\L{\Lambda}
\def\k{\kappa}
\def\f{\phi}

\def\m{\mu}
\def\n{\nu}
\def\o{\omega}

\def\p{\pi}
\def\r{\rho}

\def\x{\xi}
\def\s{\sigma}

\def\t{\tau}

\def\sF{{{ F}\!\!\!\!\hskip.8pt\hbox{\raise1pt\hbox{/}}\,}}
\def\som{{{ \omega}\!\!\!\!\hskip.8pt\hbox{\raise1pt\hbox{/}}\,}}
\def\sJ{{{\rm J}\!\!\!\!\hskip.8pt\hbox{\raise1pt\hbox{/}}\,}}

\def\F{\Phi}
\def\pa{\partial}

\def\to{\rightarrow}
\def\nonu{\nonumber \\{}}
\def\half{{1 \over 2}}

\title{Microstate solutions from black hole deconstruction}
\author[a]{Joris Raeymaekers,}
\author[b]{Dieter Van den Bleeken}

\affiliation[a]{Institute of Physics of the ASCR, \\
Na Slovance 2, 182 21 Prague 8, Czech Republic.}
\affiliation[b]{Physics Department, Bo\u{g}azi\c{c}i University\\
 34342 Bebek / Istanbul, Turkey.}

\emailAdd{joris@fzu.cz}
\emailAdd{dieter.van@boun.edu.tr}

\abstract{We present a new family of asymptotic $AdS_3\times S^2$ solutions to eleven dimensional supergravity compactified on a Calabi-Yau threefold. They originate from the backreaction of $S^2$-wrapped M2-branes, which  play a central role in the deconstruction proposal for the microscopic interpretation of the D4-D0 black hole entropy.
We  show that they are free of possible pathologies such as closed timelike curves and discuss their holographic interpretation. }

\begin{document}

\maketitle

\section{Introduction: the black hole deconstruction proposal}

Starting with the seminal work of Strominger and Vafa \cite{Strominger:1996sh}, string theory has proven highly successful in giving  microscopic accountings of the
Bekenstein-Hawking entropy of certain supersymmetric black holes.  Such accountings typically make optimal use of the protected nature of the entropy or index to
do the computation in a regime where gravitational backreaction is absent and the relevant degrees of freedom are weakly coupled D-brane excitations. This approach leaves unanswered
the question what the microstates evolve to in the regime where gravitational backreaction is significant. Furthermore, with the advent of AdS/CFT it became clear that
the black hole microstates correspond to states in the Hilbert space of a CFT which captures the degrees of freedom in a near-horizon AdS throat region.
According to the  standard AdS/CFT prescription, states in the CFT correspond semiclassically   to turning on normalizeable fluctuations of the bulk fields near the boundary, and these are expected
to lead  to  solutions of the full string/M theory on the AdS background.

Efforts to construct such  solutions within the supergravity approximation to string/M theory can be grouped loosely under the fuzzball or microstate geometry program (see \cite{Mathur:2005zp} and \cite{Bena:2013dka} for reviews and further references), although
to which extent and under which circumstances the 2-derivative low energy supergravity approximation is sufficient
for this purpose is still a matter of debate.
In this work we will make progress towards constructing supergravity solutions carrying the same charges as a large  black hole in the context of the black hole deconstruction proposal \cite{Denef:2007yt}. In this proposal, it is argued that the leading contribution to the entropy of a 4D black hole  arises from the large degeneracy of states carried by certain wrapped M2-branes, which
so far were approximated  as probes in the background of other rigid constituent branes. Our goal in this work is to construct the fully backreacted solutions\footnote{A first attempt in this direction appeared in \cite{Levi:2009az}, and we will  comment in detail on the relation with present work below.}. Our solutions contain brane sources near which the supergravity approximation breaks down, as might have been expected.  Following the terminology of \cite{Bena:2013dka} we will refer to such solutions as microstate solutions as opposed to smooth microstate geometries.

 Let us  briefly review the main ingredients of the black hole deconstruction proposal. We start from the setup first introduced and studied by
 Maldacena, Strominger and Witten (MSW) \cite{Maldacena:1997de}: consider M-theory on the background $\RR^{1,3}\times S^1 \times X$, with $X$ a Calabi-Yau threefold. When the radius of the circle is small in 11D Planck units, the type IIA string theory picture is appropriate. One can consider BPS states which are point-like in  $\RR^{1,3}$, arising from wrapped (D6, D4, D2, D0) branes\footnote{The D2/M2 brane charges discussed here, denoted as $q_A$, should not be confused with the D2/M2 charge we introduce at a later stage, which we will denote by $q_\star$. The first type correspond to 2-branes fully spatially wrapped inside the Calabi-Yau, while the second type corresponds to 2-branes fully spatially extended in the 4/5 external dimensions. } and labelled by a charge
 vector $\Gamma=(p^0, p^A, q_A, q_0)$. In the M-theory frame, these lift to (KK monopole, M5, M2,  momentum) charges respectively, but we choose to use the IIA language throughout this paper. It is possible to construct a regular black hole carrying
 D4-D0 charges $(0, p^A, 0, q_0)$ which breaks half of the supersymmetry of the background\footnote{For simplicity, we will not consider the effect of adding $D2$ charge $q_A$.} and whose Bekenstein-Hawking entropy can be computed to be:
 \begin{equation}
 S = 2\p \sqrt {q_0 p^3 }\label{D4D0entropy}
 \end{equation}
where $p^3 \equiv D_{ABC} p^A p^B p^C$  where  is triple self-intersection of the four-cycle in $X$ wrapped by the D4-brane.

We then proceed to take an M-theory  decoupling limit
\be
{R \over l_{11}} \to \infty\,,\qquad  V_\infty \equiv {V_X\over l_{11}^6}\quad\mbox{fixed,}\label{decoupling}
\ee
where $R$ is the radius of $S^1$ and $l_{11}$ the 11D Planck length. For a more detailed discussion of this decoupling limit see \cite{deBoer:2008fk}. Note that one can define a 't Hooft like coupling that is invariant under this limit:
\be
\l \equiv {p^3   \over V_\infty},\label{defthooft}
\ee

When this parameter is large, $\l \gg 1$, the bulk theory, M-theory in a  (locally) $AdS_3 \times S^2 \times X$ attractor throat geometry, is well described by its supergravity approximation as the curvature radius of $AdS_3$ and $S^2$ is $l=\l l_{11}$. To be precise the decoupled near horizon geometry originating from the 4d black hole/5d black string is not  global $AdS_3 $ but  rather a BTZ black hole, $\mathrm{BTZ} \times S^2 \times X$.

When on the other hand $\l \ll 1$ the theory is more naturally described as the low energy limit of the M5-brane worldvolume theory dimensionally reduced over the Calabi-Yau directions to a 1+1 dimensional sigma model \cite{Maldacena:1997de, Minasian:1999qn}. This incompletely understood theory is referred to as the MSW CFT. It has, up to terms subleading in the $p^A$, central charges $c_L=c_R = p^3$, and possesses $(4,0)$ superconformal symmetry. In terms of the conformal generators, the D0-charge corresponds to $q_0 = \bar L_0 - L_0$, so that the  BPS states which  contribute to the black hole entropy take the form of  Ramond ground states in the left-moving sector tensored with  highly excited states on the right-moving side. They can be easily counted in the Cardy regime ${\bar L_0 \over c} \sim {q_0 \over p^3} \gg 1$, which is also the regime where the BTZ black hole has a large horizon, and their exponential degeneracy correctly reproduces the Bekenstein-Hawking entropy \eqref{D4D0entropy} of the original 4D black hole \cite{Maldacena:1997de}.

\begin{figure}\begin{center}
\includegraphics[width=100pt]{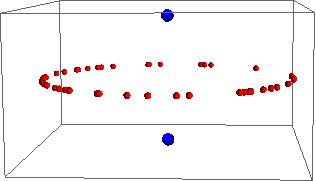}\qquad
\includegraphics[height=90pt]{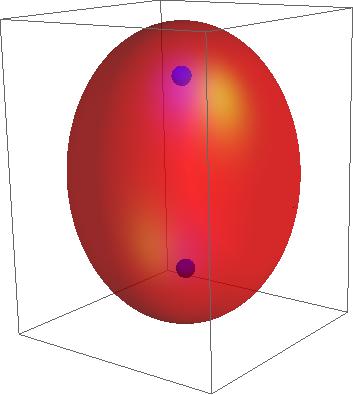}\qquad
\includegraphics[height=100pt]{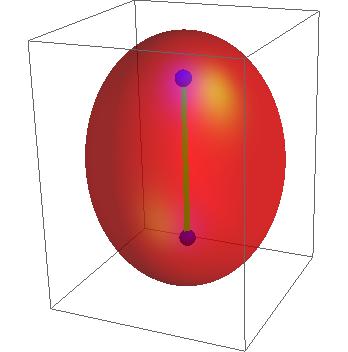}\\
(a)\hspace{3.5cm} (b)\hspace{3.5cm} (c)
\end{center}\caption{Different ways of adding D0-brane charge (in red) to a two-centered D6-anti-D6 system (the blue dots). (a) In the form of
separate D0-brane centers. (b) In the form of an ellipsoidal D2-brane with worldvolume flux. (c) Tadpole cancellation requires adding a fundamental string running between the D6 and anti-D6 branes, shown here as a green line.}\label{fancyfig}
\end{figure}

 The black hole deconstruction proposal \cite{Denef:2007yt} gives a tentative description of the typical microstates in the gravity regime $\l \gg 1$, as a particular bound state of low-entropy D-brane centers.
 One starts with a two-center D6-anti-D6 configuration with worldvolume fluxes turned on, carrying the following charges:
 \begin{align}
 &\G_{D6} = \left(1,{p^A \over 2}, {D_{ABC}p^B p^C\over 8}, - {p^3 \over 48}\right)\nonu
 & \G_{\overline{D6}} = \left(-1,{p^A \over 2}, -{D_{ABC}p^Bp^C\over 8}, - {p^3 \over 48}\right). \label{d6ads6flux}
 \end{align}
 The corresponding two-center  supergravity solution can be constructed using the methods of \cite{Bates:2003vx}, and upon taking the decoupling limit (\ref{decoupling}), one obtains the global $AdS_3$ geometry of the form $AdS_3 \times_{rot} S^2 \times X$, where the subscript $_{rot}$ means the $S^2$ is nontrivially fibered. As we will review below, this solution represents,
 in a semiclassical sense, the Ramond ground state with maximum R-charge in the MSW CFT. To obtain a solution carrying the same charges as the 4D black hole we have to add to the system an extra D0 charge $q_0  + {p^3 \over 24}$. One way to add this  charge is in the form of many separate D0-brane centers, localized in the plane between the D6 and anti-D6. Such solutions can also be constructed using the methods of \cite{Bates:2003vx} and are illustrated in figure \ref{fancyfig} (a).
 However, despite there being a large moduli space of such solutions, the discrete set of states obtained upon quantization does not account for a sizeable fraction  of the black hole entropy \cite{deBoer:2008zn}. Another way to add the D0-brane charge is, in the spirit of the Myers effect \cite{Myers:1999ps}, in the form of a D2-brane with worldvolume flux, which can supersymmetrically  wrap an ellipsoid
 with the D6 and anti-D6 branes at its centers, see figure \ref{fancyfig} (b). What makes these configurations relevant for the black hole entropy is that they couple to the D4-brane flux on the Calabi-Yau through the worldvolume Wess-Zumino coupling $\int C_3$. Due to this coupling the D2-brane  behaves as  a particle in the magnetic fields threading the  Calabi-Yau space, and has lowest Landau level degeneracy proportional to $p^3$. The combinatorics of distributing the total D0-charge over such D2-brane configurations then correctly accounts for the black hole entropy \cite{Gaiotto:2004ij}.
In the decoupling limit (\ref{decoupling}), this D2-brane configuration becomes an M2-brane which wraps the $S^2$ and is  point-like in the $AdS_3$ part of the geometry. As we shall review below, this brane traces out a  helical curve in $AdS_3$ whose radius is related to the  D0-charge.

Although the configurations (a) have long been established as fully backreacted supergravity solutions, the configurations (b) were only constructed as M2 probes in a supergravity background. The goal of this work is to progress beyond the probe approximation for this wrapped M2-brane and construct the fully backreacted geometry. In doing so  we will find that it is free  of pathologies such as closed timelike curves, which plagued our earlier attempt in this direction \cite{Levi:2009az}, and has a standard
 asymptotically $AdS_3$ behaviour consistent with expectations from the MSW CFT.

 We should also mention one complication that we will not address in this work, which arises from a worldvolume tadpole on the D6 branes of the type  discussed in \cite{Brodie:2000yz}.  The  D2 brane surrounding the D6-anti-D6 system produces a magnetic 6-form flux $F_6$ which induces a tadpole on the compact D6 worldvolume through the Wess-Zumino coupling $\int A\wedge F_6$. This tadpole can be cancelled by letting a fundamental string run between the D6 and anti-D6 branes, see figure \ref{fancyfig}(c). Furthermore, it can be argued that this string also carries an anti-D2 charge, so that the net D2-charge of the full configuration is zero\footnote{A simple argument (for which we thank F. Denef) goes as follows: Start from a D2 brane extended in the external 4D space, which does not envelop the D6 and anti-D6 centers. Such a brane wraps a contractible cycle  and carries no net D2-charge. Expanding it so as to envelop the D6 and anti-D6 centers one obtains the D2-F1 configuration of interest whose net D2 charge must still vanish.}. The M-theory decoupling limit of this configuration includes and additional anti-M2 brane at the center of AdS.
 Ignoring this tadpole does not lead to a direct inconsistency in the 5D supergravity picture we will use, as it is an effect in the internal Calabi-Yau directions.
 Nevertheless one might worry that not cancelling it leads to solutions which are ill-behaved in some way.  We will find that  this is not the case, and that the main effect of  ignoring it  is, as far we can see,  that the boundary theory is deformed by   source terms proportional
  to the M2-charge, which otherwise would be absent.

 This paper is organized as follows. In section \ref{sec3D} we review how the problem can be effectively reduced to a three dimensional description, a picture we will use in most of the article. We then discuss some physical properties of M2-brane probe particles, which originate from wrapping the internal $S^2$, in section \ref{secprobe}. The main new contributions of our work can be found in sections \ref{secM2center} and \ref{secM2spiral}. First we work out the details of the backreacted solution for an M2 at the center of $AdS_3$ and discuss at length various physical and holographic properties of this solution. In section \ref{secM2spiral} we present an additional family of solutions that tentatively describe the M2 particles spiralling at finite radius in $AdS_3$. We then connect back to the original 5d setup in section \ref{sec5D} where we
 also discuss the supersymmetry properties of the solutions. We conclude in section \ref{secout} with a short outlook on possible future directions. For the convenience of the reader we also included the appendices \ref{appnb}, \ref{apphol}, \ref{app5D} and \ref{appKS} containing various technical details.

\section{Effective three-dimensional description}\label{sec3D}
As explained in the Introduction, the brane configurations  we are interested in can be described as supersymmetric excitations of the long wavelength approximation to M-theory on the background $AdS_3 \times_{rot} S^2 \times X$, arising from wrapping an M2-brane on the $S^2$. We will make the approximation that the M2-brane charge is smeared on $X$, so that  we can construct our solutions, after dimensional reduction on $X$, within  5D supergravity  or, upon further reduction  on $S^2$, in a three dimensional theory.
We will use the simpler 3D point of view in most of the paper, and will comment on the geometric structure of our solutions from the 5D point of view in section \ref{sec5D}.

As we will see in more detail below, the M2-brane provides a source for  the volume modulus of $X$, which we will call the dilaton $\t_2$, as well as for an axion $\t_1$ which is obtained from dualizing the M-theory three-form with all legs in the 5D part of the geometry. We will often combine these in a complex field $\t = \t_1 + i \t_2$, which we will refer to as
the  axion-dilaton since it parametrizes the coset $SU(1,1)/U(1)$ just like the familiar axion-dilaton of type IIB  supergravity/string theory.  It was shown in \cite{Levi:2009az}, to which we refer for more details and conventions, that the  consistent 11D reduction
ansatz for  our solutions is
\bea
ds^2_{11}&=&\tilde \tau_2^{-2/3}\left(ds^2_3+\frac{l^2}{4}\left( d\theta^2 + \sin^2 \theta (d\f -\cala)^2\right)\right)+l_{11}^2\tau_2^{1/3}d s^2_{X}\,, \label{11Dreduction}
\eea
Where $\tilde \t_2 = {\t_2 \over V_\infty}$ denotes the fluctuating part of the dilaton field
 and the Calabi-Yau metric $d s^2_{X}$ is
assumed to be normalized to have unit volume.
The U(1) gauge field $\cala$ incorporates the possibility of having a nontrivially fibered $S^2$.

The metric above (together with an appropriate 3-form) is a solution to 11D supergravity when the effective 3D fields parameterizing it extremize the action
\be
S = {1 \over 16 \p G_3} \int_\calm \left[ d^3 x \sqrt{-g} \left( R + {2 \over l^2} - {\pa_\m \t \pa^\m \bar \t \over 2 \t_2^2 }\right)+ {l \over 2}  \cala\wedge d \cala \right]\label{3daction}
\ee
where
\be 16\p G_3 = {l_{11}^3 \over 2 \p^2 V_\infty l^2} .\ee
This is equivalent to solving the equations of motion
\begin{align}
&R_{\m\n} + {2\over l^2} g_{\m\n} - {\pa_{(\m} \t \pa_{\n)} \bar \t \over  2\t_2^2} = 0\label{metriceq1}\\
&\Box \t + i { \pa_\m\t\pa^\m\t \over \t_2} =0\label{taueq1}\\
&d\cala = 0.\label{Aeq1}
\end{align}

We now describe our ansatz for the  3D fields describing the solutions of interest,  which was proposed in \cite{Levi:2009az} and further justified in \cite{Raeymaekers:2014bqa}.
We assume the metric to be  stationary   and  written as a timelike fibration over a two-dimensional base, which we cover with a complex coordinate $z$. We will parameterize the metric as
\bea
ds^2 &=& {l^2 \over 4}\left[-\left( dt + \chi \right)^2 + \t_2 e^{-2\F} dz d\bar z\right]\label{ansatz}
\eea
The real field  $\F$, the one-form  $\chi$ and the axion-dilaton $\t$ are assumed to be time independent.

The equation (\ref{taueq1}) then reduces to
\be
\pa_z\pa_{\bar z} \t + i {\pa_z \t \pa_{\bar z} \t\over \t_2} = 0
\ee
This equation  allows for solutions where the axion-dilaton $\t$ is holomorphic,
\be
\t = \t (z)
\ee
which motivated our choice of ansatz. Such holomorphic solutions are naturally expected to be supersymmetric, and we will show that this is indeed the case, though we defer a detailed discussion of the supersymmetry properties of our solutions to section \ref{sec5D}.
Our ansatz can be seen as a straightforward  generalization of that in   \cite{Greene:1989ya}, which describes a codimension one BPS object in flat spacetime, to the case with a negative cosmological constant.

Choosing $\t$ to be holomorphic and substituting this in the metric equation (\ref{metriceq1}) leads to the following equations for $\chi$ and $\F$:
\bea
4\pa_z \pa_{\bar z}\F + \t_2 e^{-2\F } &=& 0\label{eqshol1}\\
d \chi + {i \over 2 } \t_2 e^{-2 \F} dz \wedge d \bar z &=&0 \label{eqshol2}
\eea
The first equation is a deformation of the Liouville equation, to which it reduces when $\t$ is constant. Given a solution to the first equation,
the second equation can be solved uniquely up to the choice of a closed one-form $\L$:
\be
\chi = 2 \Im m \pa \F + \L,
\ee
where $\pa$ is the standard holomorphic Dolbeault operator.
Finally the solution is completed by choosing a flat gauge connection $\cala$.

Note that our discussion was completely local so far. Below we will add susy-compatible  source terms, which describe  the M2-brane wrapped on $S^2$, completing the solution globally. Indeed, from the three dimensional point of view this brane looks like a  charged point particle, and will create delta-function singularities in the fields which we will examine in section \ref{secM2center}.

Before doing so, let us first discuss the solution  which describes the background to which we want to add the M2-brane. As discussed in the Introduction, this background is the decoupling limit of the fluxed D6-anti-D6 configuration (\ref{d6ads6flux}), which was worked out in \cite{Denef:2007yt},\cite{deBoer:2008fk}.
It can be found as a solution of the 3D theory (\ref{3daction}), with the AdS radius $l$ given in terms of the D4-charges as
 \be
 l = \left( {p^3 \over 6 V_\infty} \right)^{1\over 3} {l_{11}\over 2 \p}.
 \ee
The fluxed D6-anti-D6 configuration is realized as a particular solution   within the  ansatz (\ref{ansatz}) with constant axion-dilaton :
\bea
\t &=& i V_\infty\\
\F &=& \ln {\sqrt{V_\infty} (1- z \bar z) \over 2}\\
\chi &=& 2 \Im m \pa \F = - 2{z \bar z \over 1- z \bar z} d \arg z\\
\cala &=& d t + d \arg z \label{d6ad6}
\eea
The resulting  metric is completely regular and corresponds to global $AdS_3$; presented as a timelike fibration over the Poincar\'e disc $|z|< 1$.  The following  coordinate transformation takes us to standard global coordinates:
\bea
|z| &=& \tanh \r\\
t &=& 2 T \\
\arg z  &=& \a - T\label{toglobal}
\eea
and we obtain
\be
 ds^2 = l^2 \left[-\cosh^2 \r dT^2 +  d\r^2 + \sinh^2 \r d\a^2\right] .\label{globalads}
\ee
The Wilson line for the $U(1)$ gauge field $\cala$ means that the $S^2$ is nontrivially fibered, so that the full 11-D geometry is of the form  $AdS_3 \times_{rot} S^2 \times X$ as
anticipated in the Introduction. Its first effect is to break the symmetry\footnote{Invariance for the gauge field means that under an isometry it transforms by a corresponding gauge transformation that is well behaved on the boundary:
$\call_{K}\cala=d\lambda\,,\ \lim_{\rho\rightarrow \infty}\lambda\rightarrow \mathrm{cst}$.
} to $U(1)_L \times SL(2,\RR )_R$. Furthermore this Wilson line is singular at  $\r=0$  and has the effect of changing the periodicity of the fermions when encircling the center   of $AdS_3$.
 As we will review below, the  interpretation in the dual (4,0) theory is that it represents the Ramond ground state with maximal R-charge on the left-moving side and the $sl(2)$ invariant vacuum on the right-moving side. In our conventions, this  state has conformal dimensions $(h, \bar h) = (0, -c/24 )$ and R-charge  $j = c/12$.

We can also consider the more general class of solutions obtained by shifting both $\chi$ and $\cala$ by $(\m - 1) d \arg z $ , leading to
\bea
\t &=& i V_\infty\label{d6ad6mu1} \\
\F &=& \ln {\sqrt{V_\infty} (1- z \bar z) \over 2}\\
\chi &=&  \left( \m - 1 - 2{z \bar z \over 1- z \bar z} \right)d \arg z\\
\cala &=& d t + \m d \arg z \label{d6ad6mu2}
\eea
The parameter $\m$ is a constant whose physical values correspond to the range $0 < \m \leq 1$.
For $\m <1$, this introduces a Dirac string singularity in $\chi$ and we obtain a singular metric with the geometry of a spinning conical defect.  These solutions represent  Ramond ground states with lower than maximal R-charge on the left-moving side and the $sl(2)$ invariant vacuum on the right-moving side,  with quantum numbers $(h, \bar h) = (0, -c  /24 )$ and  $j = c \m/12$.
 They can be viewed as the result of backreacting a heavy BPS particle in the center of the $AdS_3$ solution (\ref{d6ad6}) \cite{Maldacena:2000dr}.

\section{Probe approximation}\label{secprobe}
We now turn to the issue  of adding an  M2-brane  wrapped on the $S^2$ in the $AdS_3 \times_{rot} S^2 \times X$ background described by (\ref{d6ad6}).
From the 3D point of view this M2-brane is  a charged point particle, which we will refer to as the `M2-particle' in what follows.
To start with we will review some results from treating the M2-particle as a  probe \cite{Simons:2004nm,Gaiotto:2004pc,Gaiotto:2004ij,Denef:2007yt} i.e. ignoring its backreaction on the geometry.

By dimensionally reducing the M2-brane action over the $S^2$ one obtains the following 3D action for an M2-particle of charge\footnote{The factor $2\p $  is  introduced for convenience in order to reduce the number of $2\p $ factors in what follows. In the quantum theory, our ${q_\star}$ is quantized in units of $(2\p)^{-1}$.} $2 \p {q_\star}$
\be
S_{M2} = {1 \over 16 \p G_3}\left[  - 2 \p {q_\star}\int_{\calw} d\x {\sqrt{-^*g} \over \t_2}\right] + 2 \p {q_\star} \int_{\calw} A.\label{sourceterms}
\ee
where $\calw$ denotes  worldline of the M2-particle
and $A$ is the U(1) gauge field dual to the axion $\t_1$.

\begin{figure}\begin{center}
 \begin{picture}(250,100)
\put(20,0){\includegraphics[height=100pt]{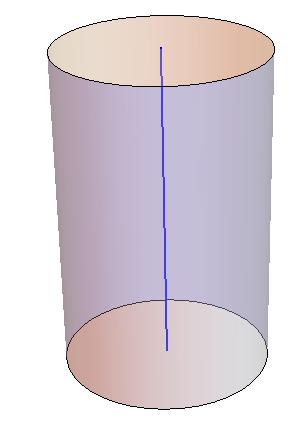}}
\put(130,0){\includegraphics[height=100pt]{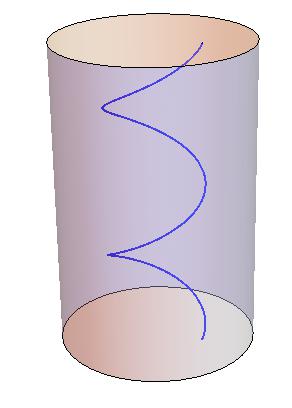}}
\put(50,-10){(a)}\put(160,-10){(b)}
\end{picture}\end{center}\caption{Trajectories of M2-particles in global AdS$_3$ depicted as a solid cylinder.
(a): an M2-particle in the `center' $\r_0=0$. (b): for $\r_0>0$ the particle traces out a helical curve.}\label{helix}
\end{figure}

The M2-particle action  (\ref{sourceterms}) in the $AdS_3$ background (\ref{globalads}) reads, in a static gauge with respect to $T$,
\be
S_{M2} =- {{q_\star} l \over 8 V_\infty  G_3} \int d T \sqrt{ \cosh^2\r- \dot{\r}^2 - \sinh^2\r \dot{\a}^2 }.
\ee
It's easy to see that a solution is provided by having the particle rotate on a helical curve at fixed $\r$ (see Figure \ref{helix}):
\be
\r = \r_0, \qquad \a =\a_0 + T \label{probetraj}
\ee
The constant  $\a_0$ can be absorbed in  $T$ and we will do so in what follows.
Note that the M2-particle is static with respect to the time coordinate $t$ in the coordinate system (\ref{toglobal}), in terms of which the metric is not manifestly static.

Let us comment on the symmetry properties of this class of solutions. It's useful to
represent  points in $AdS_3$ as $SL(2,\RR )$ group elements
\be
g(T,\r ,\a) = \left(
\begin{array}{cc}
 X_1 + X_4 & -X_2-X_3  \\
 X_2-X_3  & X_1-X_4  \\
\end{array}
\right); \qquad X_1^2 + X_2^2 -X_3^2-X_4^2 =1
\ee
with
\bea
X_1 &=& \cos T \cosh \rho \qquad X_3 = \cos \alpha \sinh \rho\\
X_2 &=&\sin T \cosh \rho \qquad X_4= \sin \alpha  \sinh \rho.
\eea
The worldlines  of the  solutions (\ref{probetraj}) are represented by
\be
g_{wl} (\r_0; T) = g(T,\r_0, T).
\ee
Each of these M2-particle worldlines preserves a specific $U(1)_L \times U(1)_R$ subgroup of the $SL(2,\RR )_L \times SL(2,\RR )_R$ isometry group
of the $AdS_3$ metric. This is easy to see for the worldline with $\r_0=0$, which is invariant under translations of both $T$ and $\a$.
The M2-particle worldlines with $\r_0 >0$  are simply related to this one by the action  of  the broken generators in $SL(2,\RR)_R$:
\bea
g_{wl} (\r_0; T) &=& g_{wl} (0; T)\cdot R(\r_0) ;\\
R(\r_0) &=& \left(
\begin{array}{cc}
 \cosh \r_0  & \sinh \r_0  \\
 \sinh \r_0  & \cosh \r_0  \\
\end{array}
\right)\label{finiterho}
\eea
Each of the worldlines (\ref{probetraj}) therefore preserves a $U(1)_L \times U(1)_R$ whose embedding in $SL(2,\RR )_L \times SL(2,\RR )_R$ depends on $\r_0$.

Another way to state this is that we can generate the M2-particle at finite radius $\r_0$ from the one at $\r_0$ through the coordinate transformation determined by
\be
g(\tilde T, \tilde \r , \tilde \a ) =  g( T,  \r ,  \a )\cdot R(\r_0 )
\ee
More explicitly it is given by
\begin{eqnarray}
\cosh^2\tilde \rho&=&\cosh^2(\rho+\rho_0)-\sin^2\frac{\psi}{2}\sinh 2\rho\sinh 2\rho_0\\
\tilde x_+&=&x_+ -x_- +\arg\left[(1+e^{ix_-}\coth\rho_0 \tanh\rho)(1+e^{ix_-}\tanh\rho_0 \tanh\rho)\right]\\
\tilde x_-&=&\arg\left[(1+e^{ix_- }\coth\rho_0\tanh\rho)(1+e^{-ix_-}\tanh\rho_0\tanh\rho)\right]\label{coordtransforho0}
\end{eqnarray}
Near the $AdS$ boundary $\r \to \infty$ this transformation takes the form
\bea
\tilde x_+ &=& x_+ + \calo (e^{- 2 \r} )\\
e^{i \tilde x_-} &=& { \cosh  \r_0 e^{i x_-} +\sinh \r_0 \over \sinh \r_0 e^{i x_-} + \cosh \r_0 } +\calo (e^{-2 \r} )\label{nbytransfo}
\eea
We see that this reduces on the boundary to a purely right-moving conformal transformation in the $SL(2,\RR )$ subgroup of the conformal group, which in terms of Virasoro generators can be written as
\be
e^{\r_0 (\tilde L_{-1} - \tilde L_1)}.
\ee

 As was shown in \cite{Simons:2004nm,Denef:2007yt,Levi:2009az} from a worldvolume $\k$-symmetry analysis,  the M2-particle solutions (\ref{probetraj}) are BPS, preserving half of the supersymmetry of the background.  Furthermore, the solutions with different values of $\r_0$ are mutually supersymmetric. This can also be understood from the point of view of the asymptotic superalgebra, since  the solutions with different $\r_0$  are related by a purely right-moving conformal transformation on the boundary,  which does not affect the asymptotic  supercharges which reside in the left-moving sector.
These observations will prove useful to obtain a proposal for the backreacted solutions with $\r_0>0$ from the one with $\r_0 =0$, as we shall see in section \ref{secM2spiral}.

We now turn to the determination of some of the worldvolume Noether charges. One easily computes the energy $H_T$ with respect to $\pa_T$  and the $\a$-angular momentum $P_\a$ of  of the solutions (\ref{probetraj}):
\bea
H_T &=&{ {q_\star} \over V_\infty} {c \over 12} \cosh^2 \r_0\\
P_\a &=& { {q_\star} \over V_\infty} {c \over 12} \sinh^2 \r_0.
\eea
where we have introduced the  Brown-Henneaux central charge
\be
c = {3 l \over 2 G_3}=p^3.
\ee
These suggest that the addition of the M2-particle changes the scaling dimensions  in the dual CFT as follows
\bea
\D h^{probe} &=&  \half ( H_T - P_\a )= {{q_\star} \over V_\infty} {c\over 24}\label{l0probe1} \\
\D \bar h^{probe} &=&  \half ( H_T + P_\a )= {{q_\star} \over V_\infty} {c\over 24} \cosh 2 \r_0.\label{l0bprobe1}
\eea
We note that  the difference of the conformal dimensions $P_\a$ becomes the D0-charge $\D q_0$ after dimensional reduction on the $\arg z$ circle.
Hence the greater the radius $\r_0$, the greater the D0-charge.
Furthermore, viewing the M2-probe probe solution from the 5D point of view, one finds  that it carries no $\f$ angular momentum $ J_\f$ on the $S^2$, which translates into a statement on the R-charge in the CFT:
\be
\D j^{probe} = J_\f =0.\label{jprobe}
\ee
Similarly one verifies that adding an M2-particle probe static with respect to $t$ to the more general backgrounds (\ref{d6ad6mu1}-\ref{d6ad6mu2}) leads to the same changes in the  quantum numbers (\ref{l0probe1},\ref{l0bprobe1},\ref{jprobe}).

In the fully backreacted solution one expects these probe predictions to be corrected due to energy and angular momentum stored in the interactions of fields sourced by the M2-particle. Naively one might expect these corrections  to appear at second and higher orders  in a perturbative expansion in the M2-brane charge ${q_\star}$. This is indeed what we will find for the right-moving scaling dimension
$\D \bar h$. On the left-moving  side, where supersymmetry resides, things will turn out to  be more subtle because of the existence of the spectral flow isomorphism of the $N=4$ superconformal algebra \cite{Schwimmer:1986mf}, which in the bulk corresponds to a coordinate redefinition which doesn't vanish near the boundary \cite{David:1999zb}. It will turn out that  the backreaction produces such a redefinition,
 which will  modify the relations (\ref{l0probe1}, \ref{jprobe})  already at linear order in ${q_\star}$. We will find in particular that in the backreacted configuration $\D h=0$ which is characteristic  for a Ramond ground state in the dual CFT.

\section{Backreacted M2-particle in the center of $AdS_3$}\label{secM2center}
In the next sections we will describe the backreaction on the 3D supergravity fields of an M2-particle moving on one of the BPS trajectories (\ref{probetraj}).
For simplicity, we will first consider the backreaction of a probe in the `center' of $AdS$, at $\r_0=0$ as in figure \ref{helix}(a), and discuss in detail its physical properties and
holographic interpretation. Having obtained this solution we will describe in section \ref{secM2spiral}
 how to act with broken symmetry generators in order to obtain solutions that thus tentatively describe the backreacted M2-particle moving on a helical curve with finite radius.

\subsection{Setting up the equations}
We will first set up the  equations following from the action (\ref{3daction}) in  the presence of the source terms (\ref{sourceterms}) produced by an M2-particle with charge
$2 \p {q_\star}$ placed at the `center'
$\r_0 =0$ of $AdS_3$. Reverting to the coordinates $t, z, \bar z$ of the ansatz (\ref{ansatz}), the M2-particle is located at $z =0$.
Varying the combined action (\ref{3daction}, \ref{sourceterms}) with respect to $\t$, one finds that
(\ref{taueq1}) gets modified to
\be
\pa_z  \pa_{\bar z} \t +  i {\pa_z \t \pa_{\bar z} \t\over \t_2} = - {  \p  i {q_\star} } \d^2 (z, \bar z)\label{tauandsource}
 \ee
It's  straightforward to check that the imaginary part of this equation follows  from  the $\t_2$ variation of the combined action (\ref{3daction}, \ref{sourceterms}); the real part requires more work as the source is written in terms of the $U(1)$ field $A$ dual to the axion. However, the real part of (\ref{tauandsource})   is guaranteed to work out by  supersymmetry, which requires $\t$ to be holomorphic.
The equation (\ref{tauandsource}) is solved by
\be
\t = - i {q_\star} \ln z + i  V_\infty \label{tausol1}
\ee
Note that, as expected, the M2-charge induces a monodromy of $\t$ when encircling the M2-particle:
\be
\t \to \t + 2 \p {q_\star} \qquad {\rm under}\  \psi \to \psi + 2 \p.
\ee
In principle, we could have added to (\ref{tausol1}) an arbitrary holomorphic function  regular in $z=0$; however since our
configuration must preserve  rotational  symmetry in the plane transverse to the M2-worldline (which is the diagonal subgroup of the $U(1)_L\times U(1)_R$ symmetry referred to in the  previous section), such additional terms are forbidden\footnote{We will find an interesting difference with the backreaction of codimension 2 axion-dilaton charged objects in asymptotically flat spacetimes which were constructed  in \cite{Greene:1989ya}. In that situation, the additional holomorphic terms are required to obtain a solution with finite energy and can be seen as introducing further sources which break the rotational symmetry \cite{Braun:2008ua}. In our asymptotically AdS case, we will find that (\ref{tausol1}), without additional  holomorphic terms,  corresponds  to an asymptotically AdS solution
with finite energy. More precisely we will argue that it corresponds to a state in a  dual field theory which is deformation of the MSW CFT.}.

Our notation $V_\infty$ for the constant term in (\ref{tausol1}) requires some explanation: as we will see below, the backreacted solution has a conformal boundary at some radius $|z_0|$. By rescaling $z$ we can
assume the boundary to be at $|z|=1$ without loss of generality, and $V_\infty$ then represents the boundary value of dilaton, which has the meaning of the size of the Calabi-Yau space in 11D Planck units. In order for
the supergravity approximation to be reliable, we will require
\be
V_\infty \gg 1 .
\ee

Now we turn to the source terms coming from varying the combined action (\ref{3daction}, \ref{sourceterms}) with respect to  the metric. Our metric ansatz (\ref{ansatz})
explicitly involves $\t$, and  careful examination shows that, if $\t$ satisfies the sourced equation (\ref{tauandsource}), the equations (\ref{eqshol1}, \ref{eqshol2}) for $\chi$ and $\F$
do not receive any delta function terms. In particular, $\chi$ should remain free of Dirac string singularities, so from (\ref{eqshol2}) the
expansion of $\F$ near the origin should not include a logarithmic term\footnote{This statement is  equivalent to the result that D7-brane sources do not produce conical singularities \cite{Bergshoeff:2006jj}.}:
 \be
 \lim_{|z|\to 0} {\F \over \ln |z|} = 0.\label{ascond1}
 \ee
The requirement of rotational invariance furthermore implies that we can choose $\F$ to be a function of $r=|z|$ alone.
To complete the solution, we also have to specify the flat connection $\cala$. Since it is not sourced by the M2-particle nor coupled to any of the fields sourced by it, we take $\cala$
to be the same as for the D6-anti-D6 solution, namely (\ref{d6ad6}).

 To simplify the equations somewhat, it will be useful to map the coordinate $z$ on the disc to  a  coordinate $w$ on the semi-infinite cylinder
 \be
 z = e^w.
 \ee
 Furthermore, we set
 \be
 w = x + i \psi
 \ee
 so the the M2-particle source is  at $x\to - \infty$ and 
 the conformal boundary at $x\to 0$.
In order for our ansatz (\ref{ansatz}) to be invariant under conformal transformations, $e^{-2 \F}$ must transform not as a scalar but as a density. Denoting  the  field in the $w$-frame by $\F^{\mathrm{cyl}}$, we have
\be
\F^{\mathrm{cyl}} (w) = \F (w) - \half (w + \bar w).
\ee
It will be convenient to to make  a  further shift
\be \tilde \F = \F^{\mathrm{cyl}} - \half \ln V_\infty\ee
which makes manifest the property that
the 3D metric actually only depends on the combination
\be
\e \equiv {{q_\star} \over V_\infty}
\ee
which is a small parameter in the regime of interest.

The field  $\tilde \F$ depends on $x$ alone due to rotational invariance, and  must satisfy the nonlinear ODE
\be
\tilde \F'' + (1 - \e x) e^{- 2\tilde  \F} =0 \label{Phieq}
\ee
Our task will be to solve this equation
subject to  (\ref{ascond1}), which in the new variables becomes the asymptotic condition
\be
\tilde \F \xlongrightarrow{x\to - \infty} - x +\calo (1).\label{xinfty}
\ee
From this behaviour we see that the solution for $\chi$ which is free of Dirac string singularities as $x\to - \infty$ is
\be
\chi = (\tilde \F' + 1) d \psi .
\ee
Furthermore, in order to assure that the solution describes the M2-particle backreacted in the background (\ref{d6ad6}), we will look for solutions that reduce
to (\ref{d6ad6}) in the limit that the M2-particle charge ${q_\star}$ is taken to zero:
\be
\lim_{\e \to 0}\tilde \F = \ln  \sinh ( - x).\label{p0AdS}
\ee
We will see that, under these conditions, we are led to a solution of (\ref{Phieq}) which is asymptotically $AdS_3$, which in terms of $\tilde \F$ means that
\be
\tilde \F  \xlongrightarrow{x\to 0_-} \ln ( - x ) +\calo (1)\label{x0bdy}
\ee
As explained above, we have chosen the coordinate $x$ such that the conformal boundary is at $x\to 0_-$.

To recapitulate, we have reduced the problem to determining a single function $\tilde \F (x)$, which has to solve (\ref{Phieq}) under the conditions (\ref{xinfty}) and (\ref{p0AdS}).
The fields of our solution are then given by
\bea
\t &=& {q_\star} \psi  + i ( V_\infty - {q_\star} x)\label{taurotinv}\\
ds_3^2 &=& {l^2\over 4} \left[ - (dt + (\tilde \F'+1) d\psi )^2 +  (1 - \e x)e^{-2 \tilde \F} (dx^2 + d\psi^2)\right]\label{metrrotinv}\\
\cala &=& dt + d\psi\label{Arotinv}
\eea

\subsection{Perturbative solution}
We now turn to the solution of (\ref{Phieq}) under the conditions (\ref{xinfty},\ref{p0AdS}). As we will explain  in more detail in section \ref{secgensol}, (\ref{Phieq}) is equivalent to a first order  Abel equation which does not belong to any subclass that has currently been solved.
It turns out however that when considering the problem as a perturbative expansion in
$\e = {{q_\star} \over V_\infty } \ll 1$ one can find an iterative solution, explicit up to quadrature, to all orders. To start we make a power series ansatz for
$\tilde \F$:
 \be
 \tilde \F = \tilde \F_0 + \sum_{\epsilon=1}^{\infty} \tilde \F_n \e^n\label{pertsolgen}
 \ee
The subsidiary condition (\ref{p0AdS}) fixes $\tilde \F_0$ to be
\be
\tilde \F_0 = \ln  \sinh ( - x).
\ee
The non-linear equation \eqref{Phieq} then decomposes order by order in $\e$ into the following linear equations
\begin{equation}
\tilde \F_n''-\frac{2}{\sinh^2 x}\tilde\Phi_n=S_n(x)\label{neqns}
\end{equation}
where
\begin{equation}
S_n(x)=\frac{1}{\sinh^2 x}\left(x\sum_{\vec{p}\in\calp_{n-1}}\prod_{l=1}^{n-1}\frac{(-2\tilde{\F}_l)^{p_l}}{p_l!}-\sum_{\vec{p}\in \calp_n^\prime}\prod_{l=1}^{n-1}\frac{(-2\tilde \Phi_l)^{p_l}}{p_l!}\right)
\end{equation}
With $\calp_n$ we denote the set of integer partitions of $n$, namely all positive integer vectors $\vec{p}$ such that $\sum_l l p_l=n$, and $\calp_n^\prime$ is that set minus the 'trivial' partition of $n$, i.e. $n$ itself, which corresponds to $\vec{p}=(0,0,\ldots,1)$.

It is important to note that $S_n$ is fully determined, and actually algebraic, in terms of the $\tilde{\F}_i$ with $i<n$, so that \eqref{neqns} can be solved iteratively.
It is  interesting that the homogeneous part of \eqref{neqns} is identical at each order, it has the two linearly independent solutions
\begin{equation}
a(x)=\coth x \qquad b(x)=1-x\coth x
\end{equation}
One can then solve \eqref{neqns} including the source term  by the method of variation of parameters. Using an argument by induction the unique solution satisfying the boundary conditions (\ref{xinfty},\ref{x0bdy}) can be found to be
\begin{equation}
\tilde{\Phi}_n(x)=(x\coth x-1)\int_{-\infty}^{x} S_n(u)\coth u\,du+\coth x\, \int_x^0 S_n(u)(u\coth u-1)du\label{allordersol}
\end{equation}
The first few orders can be integrated explicitly to yield
\bea
\tilde \F_1 &=& \frac{1}{2} (-x -x \coth x +1)\\
\tilde \F_2 &=& \frac{1}{24} \left(-6 \text{Li}_2\left(e^{2 x}\right) \coth x-3 \left(2 x^2+x^2 \text{csch}^2 x+4 \log
   \left(-2 \sinh x\right)-1\right)\right. \nonu
   && \left. +\left(\pi ^2-6 (x-2) x\right) \coth x\right)\label{pertsol}
\eea
Finally, as we will need this later, we also work out the leading terms at small $x$:
\begin{eqnarray}
\tilde{\F}_1(x)&=&-\frac{1}{2}x-\frac{1}{6}x^2+\calo(x^3)\nonumber\\
\tilde{\F}_2(x)&=&\frac{1}{6}x^2\log (-x)+\frac{1}{6}(\log 2-\frac{19}{12})x^2+\calo(x^3)\label{x2terms}\\
\tilde{\F}_n(x)&=&\left(\frac{1}{3}\int_{-\infty}^0 S_n(u)\coth u\,du\,\right) x^2+\calo(x^3)\qquad(n\geq 3)\nonumber
\end{eqnarray}

\subsection{Asymptotics}
The equation (\ref{Phieq}) also allows for a  perturbative expansion near the boundary, i.e for small $|x| \ll 1$, which will be useful to determine the asymptotic charges of our solution. We make the following ansatz in terms of a `transseries' in $x$:
\be
{e^{\tilde \F} } =  \sum_{n=0}^\infty P_n\left( u \right) {\e^{n-1} (-x)^n \over n!}, \qquad u \equiv \ln (- x)+ {C \over \e^2} \label{asansatz}
\ee
where $C$ is an integration constant\footnote{Note that the most general solution to (\ref{Phieq}) has two integration constants. As we demand the behaviour (\ref{x0bdy}) for  $x\to 0_-$, this fixes one of those two constants. This constant could easily be reinstated by making use of  the following
symmetry of (\ref{Phieq}):
\bea
x &\to& (1- \e c_0)x + c_0\\
\tilde \F &\to & \tilde \F +{3\over 2} \ln (1 - \e c_0).\label{symm}
\eea
for a constant $c_0$.} on which we will comment below.

 Substituting the ansatz (\ref{asansatz}) into (\ref{Phieq}) one finds that
 the condition that the AdS boundary is at $x\to 0_-$, see (\ref{x0bdy}), fixes the solutions for $P_0$  and $P_1$ to be
 \bea
 P_0&=&0\\
 P_1 &=& 1.
  \eea

For the equations determining the $P_k(u)$ for $k>1$ one finds  the recursion relations
 \begin{align}
& \ddot P_k + (2 k-3) \dot P_k + k(k-3) P_k =- 2 \d_{k,2} - \sum_{n=1}^{k-2}\left( \begin{array}{c} k \\ n\end{array}\right)  \left[ (2n-k) P_{k-n} P_{n+1}- \dot P_{k-n} P_{n+1} \right. \nonu & \left. + {3n-k+1\over n+1}
 P_{k-n} \dot P_{n+1} -
{ \dot  P_{k-n} \dot P_{n+1}\over n+1} + {P_{k-n} \ddot P_{n+1}\over n+1}\right]\label{Peqs}
\end{align}
where a dot means differentiation with respect to $u$. These equations determine the $P_k(u), k>1$ to be polynomials of order $[k/2]$ in $u$.
The integration constant $C$ in (\ref{asansatz}) arises because the coefficients in (\ref{Peqs}) are $u$-independent,
and one can verify that the system (\ref{Peqs}) does not give rise to further integration constants that cannot be absorbed in a redefinition of $C$.  These recursive equations
can be solved e.g. with Mathematica to rapidly obtain the $P_n$ up to high values of $n$. For our purposes  we will only need $P_2$ and $P_3$ as we will see that only they enter in the asymptotic charges. They are given by
\bea
P_2 &=& 1\\
P_3 &=& u.
\eea

The value of the integration constant $C$ in (\ref{asansatz}) appropriate for our solution  is determined by matching onto the correct delta-function sources in the deep interior of the bulk
where $x \to - \infty$, as expressed by the boundary condition (\ref{xinfty}). We will see below that $C$  enters in the expression for the asymptotic charge $\bar h$ which is not fixed by supersymmetry.
The $\epsilon$ dependence of $C$ is determined by comparing (\ref{asansatz}) with our power series in $\e$ (\ref{allordersol}), whose expansion to order $x^2$, (\ref{x2terms}), is sufficient for this purpose. One finds that
\be
C = 1- \e +  \left( \ln 2-\frac{5}{6} \right) \e^2 +2\sum_{n=3}^\infty \epsilon^n\int_0^\infty S_n(u)\coth u\,du.\label{Ceps2}
\ee

It's interesting to note that there exists a similar perturbative expansion of the equation for large $|x| \gg 1$, i.e. near the M2-particle position, which we derive in Appendix \ref{appnb}.
After imposing the near-brane behaviour (\ref{xinfty}), this expansion contains a single integration constant $D$ which we expect to be completely fixed by imposing the behaviour (\ref{x0bdy}) near $x\to 0_-$. The first terms of this expansion lead to
\be
   \tilde \F = -  x + \ln D - {\e +  (1 - \e x) \over 4 D^2 }e^{2  x} + \calo ( e^{3  x} ).\label{PhiM2}
   \ee
  For example, to order $\e^2$ we find from comparison to our perturbative expansion (\ref{pertsol}):
\be
D = \frac{1}{2}+\frac{\epsilon }{4}+\left(\frac{1}{8}-\frac{\pi ^2}{48}\right) \epsilon ^2 + \calo (\e^3 ).
\ee

\subsection{Properties of the 3D geometry}

Before turning to the holographic interpretation of our solution, we would like to discuss some of the properties of its three-dimensional Lorentzian geometry. As is to be expected,
the scalar curvature diverges near the M2-particle trajectory $x\to - \infty$ due to the source term (\ref{tauandsource}). The curvature of a generic solution of the system (\ref{metriceq1}-\ref{Aeq1}) with holomorphic $\t$ is
\be
l^2 R= -6 + {4 |\pa_z \t |^2 e^{2 \F} \over \t_2^3}.
\ee
For our specific solution one finds, using the near-M2 expansion (\ref{PhiM2}), the leading $x\to - \infty$ behaviour
\be
R \approx {4 D^2 \over \e (-x ) e^{2 x} }.
\ee
Hence we expect  higher derivative corrections to our effective action (\ref{3daction})   to be significant in the vicinity of the M2-particle. It would be interesting to investigate if those and the corresponding corrections to the probe brane action can be made more concrete using ideas of brane effective actions, along the lines of \cite{Michel:2014lva}.

Next we would like to discuss the possible issue of closed timelike curves. Even though we know the solution only in a perturbation expansion, we will  still be able to argue that it is actually free of closed timelike curves, making use of a result proven in  \cite{Raeymaekers:2011uz}. It was shown there that, for solutions with  two commuting isometries in a theory which obeys the null energy condition, as is  the case for us, closed timelike curves must be absent if the component $g_{\psi \psi}$ is positive
both in the vicinity of the symmetry axis  and the boundary.

In our case, the symmetry axis is at the location of the M2-particle $x \to - \infty$, where we find\footnote{Note that near a regular timelike symmetry axis, closed timelike curves are guaranteed to be absent by the equivalence principle  \cite{Mars:1992cm}; however since we have a curvature singularity on the axis we need to be more careful.}, using (\ref{metrrotinv}) and the expansion (\ref{PhiM2}),
\be
g_{\psi\psi} \sim {l^2 \e \over 4 D^2}(-x) e^{2 x} + \calo (e^{3 x})
\ee
which is indeed positive. Near the boundary $x \to 0_-$ we have, using the expansion (\ref{asansatz}),
\be
g_{\psi\psi} \sim {\left(1-{\e \over 2} \right) \over 2 (-x ) } + \calo ( \ln (- x))
\ee
  which is also positive since we argued that $\e$ must be small in the regime of validity.
We conclude that the requirements for  the theorem of \cite{Raeymaekers:2011uz} are satisfied and that our solution is free of closed timelike curves.

\subsection{Holographic interpretation: dual field theory}
From our results on the near-boundary behaviour of $\F$ we can we can derive the asymptotic behaviour of the metric and other fields and interpret the solution  holographically.
It will sometimes be convenient to use instead of the dilaton $\t_2$ the field
\be
\Psi \equiv - \ln \t_2
\ee
which has a canonical kinetic term.
Making the coordinate redefinitions
\bea
x &=&- \half \left(1 - {\e \over 2}\right) y\label{rhocoord} \\
t &=& \left(1 - {\e \over 2}\right) (x_+ - x_-) \label{boundcoords} \\
\psi &=& x_-
\eea
and using (\ref{taurotinv}-\ref{Arotinv}, \ref{asansatz}), the fields in our solution have the following asymptotic expansions near the boundary $y\to 0$:
\bea
ds^2_3 &=& l^2 \left[  {dy^2 \over 4 y^2} +   {g_{(0)}\over y}  + g_{(2)++} dx_+^2 + g_{(2)--} dx_-^2 +\ln y  \tilde g_{(2)--} dx_-^2 + \calo (y \ln y) \right]\label{metricFG}\\
\Psi  &=& \Psi_{(0)}+ y \Psi_{(2)}+ \calo (y^2 ) \\
\t_1 &=& \t_{1(0)} \\
\cala &=& \cala_{(0)+} dx_+ +  \cala_{(0)-} dx_-
\label{FG2} \eea
These fit in the general  Fefferman-Graham-type \cite{FG} expansions appropriate for asymptotically AdS solutions of the 3D theory (\ref{3daction}) which we review in Appendix \ref{apphol} (see
(\ref{FG1}-\ref{FG4})), where we also  perform in detail the
 holographic renormalization \cite{de Haro:2000xn} of our 3D theory (\ref{3daction}).

According to the standard AdS/CFT dictionary, the leading parts in these expansions are sources for various CFT operators: $g_{(0)}$ for the stress tensor $T$, $\Psi_{(0)}$ and $\t_{1(0)}$ for two $(h, \bar h) = (1,1)$ marginal operators $\calo_\Psi$ and $\calo_{\t_1}$ respectively. Furthermore, as reviewed in Appendix \ref{apphol},   the component $\cala_-$ of the Chern-Simons  gauge field plays the role of a source for one of the left-moving $SU(2)$ R-symmetry currents, $J^3_+$, of the MSW theory.
We find for our solution
\bea
g_{(0)} &=& dx^+ dx^-\label{bdymetr}\\
\Psi_{(0)} &=&- \ln V_\infty\label{dilprof}\\
\t_{1(0)} &=& {q_\star} x_- \label{axprof}\\
\cala_{(0)-}&=&  {\e \over 2} \label{chempot}
\eea
The first two expressions tell us that the CFT is defined on the flat cylinder with circumference\footnote{In fact, this was the reason for the making the rescaling (\ref{rhocoord}).} $2\p$, while the second specifies the point in the CFT  moduli space of marginal  deformations by $\calo_\Psi$. Both of these are unmodified by the addition of the M2-particle.  More interesting  are  the last two relations (\ref{axprof}, \ref{chempot}), which tell us that once the M2-charge ${q_\star}$ is nonzero, the dual CFT action is deformed by  source terms
for  $\calo_{\t_1}$ and  $J^3_+$:
\be
\d S_{CFT}=  - \int dx_+ dx_-  \left( \t_{1(0)} \calo_{\t_1} + \cala_{(0)-} J^3_+ \right).\label{MSWdef}
\ee
The second term in (\ref{MSWdef}) comes from the boundary term for the Chern-Simons field (\ref{Act}).  It's somewhat suprising to find such a source for the R-current,
 since it was absent before adding the M2-particle and the
Wilson line $\cala$ is is not directly sourced by it. However, we see from (\ref{boundcoords}) that adding the M2-particle induces a large coordinate transformation near the boundary which modifies the decomposition of  $\cala$ into left- and right moving pieces, which leads to the nonzero $\cala_{(0)-}$ in (\ref{chempot}).

 A remark is in order regarding the special form of the source terms  (\ref{dilprof}-\ref{chempot}) in our solution. Generic non-constant sources in the dual field theory imply that
 translational invariance and therefore also conformal invariance is broken.
 This explicit symmetry breaking  is encoded in a   Ward identity for the divergence of the stress tensor \cite{de Haro:2000xn},
which we  derived for our system from the bulk point of view in (\ref{Wardid}).
Our solution on the other hand belongs to a subclass where the sources are of the form
\be
\Psi_{(0)} ={\rm constant}, \qquad \pa_+ \t_{1(0)} =\pa_+  \cala_{(0)-}= 0.\label{specialsources}
\ee
 Since the sources are purely rightmoving and the dual operators have $h=1$, we do  expect to preserve the left-moving conformal symmetry. Deformations of this type are sometimes called null deformations and were studied in a holographic context in \cite{Costa:2010cn}.

 To be a bit more concrete, we substitute (\ref{specialsources}) in the expressions for the trace anomaly (\ref{tranom}) and the Ward identity   (\ref{Wardid}) for a  flat boundary metric and obtain
\bea
\langle T_{+-} \rangle &=& 0 \\
\pa_- \langle T_{++} \rangle &=& 0\\
\pa_+ \langle T_{--} \rangle &=& -\half \langle \calo_{\t_1} \rangle \t_{1(0)}'\label{holstress}
\eea
The second line suggests that left-moving conformal invariance is preserved. In section \ref{sec5D} below we  will  find evidence that the deformation also preserves some supersymmetry, namely half of the left-moving N=4 supersymmetry of the undeformed MSW theory.
The last equation indicates that right-moving translation invariance (and hence conformal invariance) is broken. Nevertheless, $\pa_+ \langle T_{--} \rangle $ does vanish on states where $\langle \calo_{\t_1} \rangle =0$
 which turns out to be the case for our solutions. This structure, where only one chiral sector of the CFT seems to be preserved/deformed is reminiscent of ideas of chiral \cite{Balasubramanian:2009bg} or warped \cite{Detournay:2012pc} CFTs that appeared in other studies of extremal black holes. It would be interesting to investigate if such a connection can indeed be concretely realized.

\subsection{Holographic one-point functions}\label{secholvevs}
Having determined some properties of the dual field theory in which our solution lives, we now turn to the determination of the holographic VEVs of various operators in the state encoded by our bulk solution.
As derived in Appendix \ref{apphol}, these can be read off from the  expansions (\ref{FG2}) as follows:
\begin{equation}
\langle T_{++} \rangle = {c \over 12 \p} \left(   g_{(2)++}  +{1\over 4}  \cala_{(0)+}^2\right),
\qquad \langle T_{--} \rangle = {c \over 12 \p} \left(   g_{(2)--} + \tilde  g_{(2)--} +{1\over 4}  \cala_{(0)-}^2\right)\label{bdyT}\end{equation}
\begin{equation}
\langle T_{+-} \rangle = 0,\qquad  \langle J^3_+ \rangle = {c \over 24 \p} \cala_{(0)+}\qquad
\langle \calo_\Psi \rangle = - {c \over 12 \p} \Psi_{(2)}, \qquad
\langle \calo_{\t_1} \rangle = 0
\end{equation}
In particular we find, for our solution,
\begin{align}
 g_{(2)++} =& -{1 \over 4}\left(1 - {\e \over 2}\right)^2, &
 g_{(2)--} =&   -{1 \over 4} \left( C +{\e^2 \over 12} +\e^2 \ln \left(\half \left(1 - {\e \over 2}\right)\right)\right), &
\tilde g_{(2)--} =& -{\e^2 \over 4}\nonu
\Psi_{(2)} =& -{\e \over 2} \left(1 - {\e \over 2}\right),&
\cala_{(0)+} =&  \left(1 - {\e \over 2}\right) &&
\end{align}
leading to the zero-mode VEVs
\bea
h&=&\langle L_0 \rangle = 0\label{vevs1}\\
\bar h&=&\langle \bar L_0 \rangle = - {c \over 24} \left( C +{5  \over 6}\e^2  + \e^2 \ln \left(\half \left({1  } - {\e \over 2}\right)\right)\right)\label{barheq}\\
j&=&\langle (J^3)_0 \rangle ={c \over 12} \left(1 - {\e \over 2}\right)\\
\langle (\calo_\Psi)_0 \rangle &=& {c \e \over 12} \left(1 - {\e \over 2}\right)\\
\langle (\calo_{\t_1})_0 \rangle &=&0\label{vevs5}
\eea

Some comments are in order. It is interesting that the left-moving weight $h$ vanishes exactly to to all orders $\e$, indicating that our solution represents a Ramond sector ground state
of the leftmoving superconformal algebra. We will find
additional evidence for this interpretation from a Killing spinor  analysis in section \ref{sec5D}.
We further  note that, due to the presence of the
source term for the R-current in (\ref{MSWdef}) the R-charge $j$ is smaller than it
was in  the background. As already anticipated in section \ref{secprobe}, these charges  differ from
the naive probe computation already at first order in ${q_\star}$. This  is related to spectral flow in the superconformal algebra: we could apply a spectral flow transformation in the bulk \cite{David:1999zb} to obtain a solution whose charges match (\ref{l0probe1}, \ref{jprobe}) to first order in ${q_\star}$.

To determine the right-moving weight $\bar h$,  we substitute the value of the integration constant $C$ determined in \eqref{Ceps2}. Somewhat surprisingly the order $\e^2$ contributions to $\bar h$ cancel, while higher order corrections remain\footnote{We were able to evaluate the third order correction exactly, while numerical methods where used for the higher orders. Contrary to second order, a complete cancellation at third order does not arise. However, it seems suggestive that a partial cancellation still occurs, eliminating the rational term originating from the expansion of the logarithm in \eqref{barheq} and leaving a purely transcendental answer. It would be interesting to understand if the numerical results at higher order have a similar interpretation.}:
\be
\bar h = - {c \over 24}\left(1-\epsilon +\frac{\pi^2}{6}\epsilon^3-(0.819...)\epsilon^4+(0.621...)\epsilon^5\right)+\calo(\epsilon^6) .\label{hbar}
\ee
To first   order in $\e$ this coincides with  the probe approximation result  (\ref{l0bprobe1}) for an M2-particle in the center of AdS.

Finally let us also discuss the 1-parameter family of solutions, labeled by $\m$ with $0<\m \leq 1$, obtained from the one above   by shifting both $\chi$ and $\cala$ by the same harmonic form $(\m-1) d\psi$, leading to
\bea
\t &=& {q_\star} \psi  + i ( V_\infty - {q_\star} x)\label{taurotinvmu}\\
ds_3^2 &=& {l^2\over 4} \left[ - (dt + (\tilde \F'+\m) d\psi )^2 +  (1 - \e x)e^{-2 \tilde \F} (dx^2 + d\psi^2)\right]\label{metrrotinvmu}\\
\cala &=& dt +\m d\psi\label{Arotinvmu}
\eea
while the solution for $\tilde \F$ remains unmodified.
Following our comments at the end of section \ref{sec3D}, we propose that these solutions represent the backreaction of  an M2-particle in the backgrounds  which correspond to   Ramond ground states with less than maximal R-charge.
The transformation to Fefferman-Graham coordinates $(y, x_+, x_-)$ now reads
\bea
x &=&- \half \left(\m - {\e \over 2}\right) y \\
t &=& \left(\m - {\e \over 2}\right) (x_+ - x_-)  \\
\psi &=& x_-.\label{FGcoordsmu}
\eea
A similar asymptotic analysis yields the result that the sources (\ref{bdymetr}-\ref{chempot}) are unmodified, hence the solutions for different values of $\m$ represent states in the same boundary theory. The operator VEVs (\ref{vevs1}-\ref{vevs5}) on the other hand change to
 \bea
h &=& 0\label{muvevs1}\\
\bar h &=& - {c \over 24} \left( C +{5  \over 6}\e^2  + \e^2 \ln \left(\half \left({\m  } - {\e \over 2}\right)\right)\right)\\
j &=&{c \over 12} \left({\m  } - {\e \over 2}\right)\\
\langle (\calo_\Psi)_0 \rangle &=& {c \e \over 12} \left({\m  } - {\e \over 2}\right)\\
\langle (\calo_{\t_1})_0 \rangle &=&0.\label{muvevs5}
\eea
Specifically, we find for the right-moving dimension
\be
\bar h = \bar h_{\m =1} - {c \e^2 \over 24}\ln {\m - {\e \over 2}\over 1 - {\e \over 2}}\label{hbarmu}
\ee
where $\bar h_{\m =1}$ is the expression given in (\ref{hbar}).
In particular, we see that for $\m\neq 1$, the order $\e^2$ contribution to $\bar h$ no longer vanishes.

\subsection{More general solutions}\label{secgensol}
We end this section with some observations on the  general solution to the equation (\ref{Phieq}), not necessarily obeying the conditions (\ref{xinfty}, \ref{p0AdS}) required to describe a backreacted M2-particle. We will in particular comment on the interpretation of the solution describing the 3D G\"{o}del universe which was was found in \cite{Levi:2009az}. This subsection can easily be skipped by the reader mainly interested in the backreacted M2-particle solutions.

We begin by changing variables from $\tilde \F (x)$  to $X(s)$ defined by 
\bea
1-\e x &=& e^{-\e s}\\
X &=& -\left( 2\tilde \F + 3 \e s + \ln{3\over 2 }\right)
\eea
Note that this transformation has a well-defined $\e\to 0 $ limit. 
 In terms of these variables (\ref{Phieq}) becomes an  autonomous (i.e. with $s$-independent coefficients) equation for $X$:
\be
\ddot{X} + \e \dot{X} + 3 ( \e^2 - e^X)=0 . \label{Xeq}
\ee
where the dot means differentiation with respect to $s$. The translation symmetry in $s$ in this form corresponds to the symmetry (\ref{symm}) in the old variables.

Before continuing the analysis of (\ref{Xeq}) we note that we can also bring (\ref{Xeq}) to a (non-autonomous) first order form by taking the dependent variable to be $X$ and the independent variable to be $A(X) = \dot X$. The equation (\ref{Xeq}) then becomes
\be
A {d A \over d X} + \e A = 3 ( e^X - \e^2).
\ee
This is a particular case of  Abel's equation of the second form. We have verified, using the techniques of \cite{Cheb-Terrab}, that it does not belong to a subclass which has been previously solved.

Returning to the autonomous form (\ref{Xeq}) of our equation, and defining $Y = \dot{X}$, the solutions to (\ref{Xeq}) can be pictured  as  flows in the $(X,Y)$ plane
\be
(\dot{X}, \dot{Y} ) = (Y, -\e Y + 3(e^X - \e^2))\label{floweq}.
\ee
\begin{figure}\begin{center}
 \begin{picture}(240,220)
\put(0,0){\includegraphics[width=240pt]{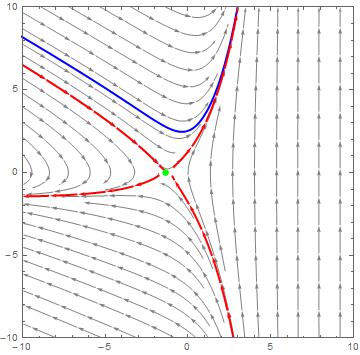}}
\put(120,200){I}
\put(200,150){II}
\put(100,30){III}
\put(30,150){IV}
\end{picture}\end{center}\caption{Phase diagram for $\e =0.5$. The blue flow line corresponds to the backreacted M2-particle, while the fixed point indicated by the green dot corresponds to the 3D G\"odel universe.}\label{flows}
\end{figure}
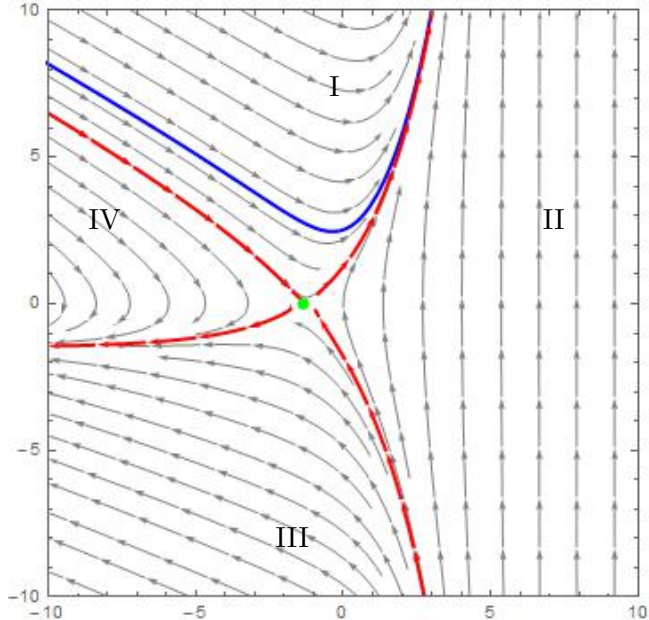
The resulting flow diagram is shown in figure \ref{flows}. All of these flows will turn out to correspond to supersymmetric solutions. The flow equation (\ref{floweq}) has a fixed point at
\be
X_G = \ln \e^2, \qquad Y_G =0\label{godsol}.
\ee
From looking at small fluctuations around the fixed point we see that there is one attractive and one repulsive direction. The tuned flows which begin or end precisely at the fixed point divide the $(X,Y)$ plane into 4 sectors, and the M2-particle solution of interest corresponds to a particular flow in sector I.

Let us comment on the solution corresponding to the fixed point (\ref{godsol}).
Translated back to the original variable $\tilde \F$ it corresponds to
\be
\tilde \F_G = \half \ln {2  (1 - \e x)^{3} \over 3 \e^2}. \label{PhiGod}
\ee
This solution, which is thus far the only known analytic solution to the equation (\ref{Phieq}), gives rise to the metric 3D G\"odel universe  and was studied in \cite{Levi:2009az}.
One of the key differences with the solutions studied in this paper is that it does {\it not} satisfy the condition (\ref{xinfty}) required to describe a single backreacted M2-particle. Nevertheless, it does allow for an interpretation in terms of (generalized) M2-particles in the following sense \cite{Raeymaekers:2014bqa}. The 3D G\"odel universe is highly symmetric, being a homogeneous space with isometry group  $U(1)_L \times SL(2,\RR )_R$, while a single
M2-particle only has the symmetry $U(1)_L \times U(1)_R$ as we argued in section \ref{secprobe}. It was proposed in \cite{Raeymaekers:2014bqa}  that the G\"odel solution instead comes from  a smeared congruence of particles obtained by
averaging  over the action of $SL(2,\RR )_R$. The resulting configuration preserves the same supersymmetry as a single M2-particle, while the bosonic symmetry is indeed
   enhanced to $U(1)_L \times SL(2,\RR )_R$.  The stress tensor of the smeared  congruence of particles is that of  pressureless rotating dust which is well-known to give
  rise to the 3D G\"odel universe.

It would be interesting to study the solutions corresponding to other sectors in the flow diagram of Figure \ref{flows}, especially those in sector II since they also have an asymptotically $AdS_3$ region.

\section{Backreacted M2-particle at finite radius}\label{secM2spiral}
We now discuss our proposal for the backreacted solution corresponding to an M2-particle moving on a helical curve at radius $\r_0$ in global $AdS_3$ as in figure \ref{helix}(b).
As  argued in section \ref{secprobe}, in the probe approximation this solution can be obtained from the one at $\r_0 =0$ by applying the coordinate transformation (\ref{coordtransforho0}).
We also  observed in (\ref{nbytransfo}) that this is a `large' coordinate transformation which on the boundary reduces to a purely rightmoving $SL(2,\mathbb{R})$ transformation.

To obtain the corresponding backreacted solution, we propose to similarly  perform a  large coordinate transformation  on the solution describing the M2-particle in the center of $AdS$. Concretely, we take our solution  (\ref{taurotinv}-\ref{Arotinv}), expressed in the coordinates $(y, x_+, x_-)$ (\ref{FGcoordsmu}), and perform the  transformation (\ref{coordtransforho0}), where $\r$ is related to $y$ as
\be
y =4 e^{-2 \r}.
\ee
 Near the boundary this coordinate transformation acts as
\bea
x_- &\to &  F(x_-) + \calo (y^2)\\
x_+   &\to & x_+ - {y \over 2} {F''(x_-) \over F'(x_-)}+ \calo (y^2)\\
y &\to &  F' (x_-) y + \calo (y^2)\label{astransfo}
\eea
where
\be
F (x_-) =- i \ln \left(  { \cosh  \r_0 e^{i x_-} +\sinh \r_0 \over \sinh \r_0 e^{i x_-} + \cosh \r_0} \right).
\ee

Applying this to  the solution (\ref{taurotinv}-\ref{Arotinv}) for the M2-particle at $\r_0=0$,  with asymptotic behaviour (\ref{metricFG}-\ref{FG2}),  we easily read off the source terms and VEVs in the new solution.
First of all, the solution represents a state in a different dual theory where the boundary sources (\ref{axprof}, \ref{chempot})  are changed   to
\bea
\t_{1(0)} &=& {q_\star} F(x_-) \label{axproftr}\\
\cala_{(0)-}&=&  {\e \over 2} F'(x_-) \label{chempottr}.
\eea

Turning our attention to the operator VEVs in the new solution, we have already observed in section \ref{secholvevs} that a coordinate transformation of the form (\ref{astransfo}) gives a new stress tensor VEV $\langle T_{--} \rangle$ which is independent of $x_+$. Explicitly we find the transformation law
\bea
\langle T_{--} \rangle &\to & (F')^2 \langle T_{--} \rangle  - {c \over 24 \p} S(F,x_-) +{c \over 12 \p} \tilde g_{(2)--} (F')^2 \ln F' \label{Ttransf}
\eea
where $S(f,x_-) $ is the  Schwarzian derivative
 \be
 S(F,x_-)  = {F'''\over F'} - {3\over 2} \left( { F''\over  F' } \right)^2.
 \ee
 The first two terms constitute the standard CFT stresstensor transformation law, while the anomalous last term
 is   due to the fact that applying the  conformal transformation gives a state in a {\em different} field theory\footnote{It's interesting to note that the anomalous
  term can be cancelled by accompanying (\ref{astransfo}) by a large gauge transformation of the form
$\cala \to \cala + d (G(x_-))$, where $G$ satisfies
$$
G' = F' \left( - \cala_{(0)-} + \sqrt{ \cala_{(0)-}^2 - 4 \tilde g_{(2)--} \ln F'} \right).
$$
This corresponds to changing the source (\ref{chempottr}) precisely such that $\langle T_{--} \rangle$ has the desired transformation law.
We will however refrain from doing this extra gauge transformation  as it would obscure some nice properties of the Killing spinors to be discussed in section \ref{sec5D}.
}. It would
 be interesting to derive the transformation (\ref{Ttransf}) from the CFT side  from the two-point function of $T_{--}$ in the deformed theory (\ref{MSWdef}).
For the specific transformation (\ref{astransfo})  applied to our solution (\ref{metricFG}-\ref{FG2}), (\ref{Ttransf})  can be worked out a bit further to give
\be
\langle T_{--} \rangle = {1\over 2 \p}\left( (F')^2 \left(\bar h_{\r_0=0} + {c\over 24}\right)-{c\over 24}- {c\e^2 \over 24}(F')^2 \ln F'\right)
\ee
where $\bar h_{\r_0=0}$ is the rightmoving  weight of the original solution (\ref{hbar}) (or (\ref{hbarmu}) for the more general solutions (\ref{taurotinv}-\ref{Arotinv})).
For the constant Fourier  mode of this expression one finds
\be
\bar h =-{c\over 24} + \cosh 2 \r_0  \left(\bar h_{\r_0=0} + {c\over 24}\right)+ {c\e^2 \over 24}\left(2 \r_0 \cosh 2 \r_0 -e^{-4\rho_0}{}_2F_1^{(0,1,0,0)}(
\frac{1}{2},2,1,1-e^{-4\rho_0})\right)
\ee
Again, to linear order in $\e$ this  coincides with the probe result (\ref{l0bprobe1}).
The only other VEV which is modified compared to (\ref{muvevs1}-\ref{muvevs5}) is
\be
\langle \calo_\Psi \rangle =  {c \over 24 \p}\e\left(1 - {\e\over 2}\right)  F' .
\ee

\section{Lift to 5 dimensions}\label{sec5D}
We will now present the uplift of  our 3D solutions to solutions of the 5D supergravity theory which arises
from dimensionally reducing  11D supergravity on a Calabi-Yau threefold. We will uncover some geometric structures
present in our solutions and will show that these are precisely of the kind required for generic solutions with nontrivial hypermultiplets
preserving at least one Killing spinor \cite{Bellorin:2006yr}.
As promised, we will also explicitly construct the full
set of Killing spinors that preserve our solutions and discuss their properties.
 The current 5D setting would also be the natural starting point for constructing the corresponding asymptotically flat solutions, which we will not
 attempt  in this work.

 Dimensionally reducing 11-dimensional supergravity on a Calabi-Yau manifold $X$ gives ungauged 5-dimensional N=1 supergravity coupled to $h_{(1,1)}-1$ vector multiplets and
 $h_{(2,1)}+1$ hypermultiplets \cite{Cadavid:1995bk}. Of these hypermultiplets, one is the universal hypermultiplet whose couplings are independent of the topology of $X$. Our solutions fit within a
 consistent truncation where the vector multiplet scalars $Y^I$ are constant while, in the hypermultiplet sector, only one of the two complex scalars within the universal hypermultiplet is allowed to vary. This complex scalar   is precisely our  axion-dilaton field $\t$. The action governing this truncation is given in (\ref{5daction}) in Appendix \ref{sec5D}, to which we also refer for more details on our conventions.

All the 3D solutions considered so far lift to 5D solutions of the following form\footnote{In this section, we adopt units in which the reduced 5D Planck length,
$\tilde l_5 \equiv {l_{11} \over 4 \p V_\infty^{1/3}}$, is set to one.}:
\bea
ds^2_5 &=& {l^2\over 4} \left[-\left( dt + 2 \Im m(\pa \F) + \L \right)^2 + \t_2 e^{-2\F} dw d\bar w + d\theta^2 + \sin^2 \theta (d\f -\cala)^2\right]\nonu
\cala &=& d t + \L, \qquad d \L =0, \qquad \t = \t(w), \qquad l =2  \left( {P^3 \over 6} \right)^{1\over 3}\\
F^I &=& {P^I \over 2} \sin \theta d\theta \wedge  (d\f - \cala), \qquad
 Y^I = {P^I \over l}  \label{5Dlift}
\eea
where $\F$ is a solution of \be 4 \pa_w \pa_{\bar w} \F + \t_2 e^{-2\F} =0 .\label{Phieqw}\ee
The metric can be rewritten in the following way as a timelike fibration over a 4D base:
\be
ds^2 = - f^2 ( dt + \xi)^2 + f^{-1} ds^2_4
\ee
where
\bea
ds^2_4 &=& - {l^2 \over 8} \cos \theta \left( \t_2e^{-2 \F} dw d\bar w + d \theta^2 + \tan^2 \theta (d \f + 2 \Im m (\pa \F) )^2 \right)\label{4Dmetric}\\
f &=& -{l \over 2} \cos \theta\\
\x &=& \tan^2 \theta d \f +  2 \sec^2 \theta  \Im m(\pa \F)   + \L
\eea

Let us first discuss the geometry of the 4D-base space. We note that it is ambipolar, changing signature as $\theta $ varies between 0 and $\p$, while nevertheless the full metric
remains  Lorentzian.
When the axion-dilaton $\t$ is constant, the base  has a hyperk\"ahler structure \cite{Gauntlett:2002nw}, which gets deformed in an interesting way for nonconstant  holomorphic $\t (w)$. We refer to refs. \cite{Bellorin:2006yr}, \cite{Raeymaekers:2014bqa} for more details on the general structure of such solutions\footnote{More specifically, our solutions are of the type of eqs. (3.54-3.58) in \cite{Raeymaekers:2014bqa}, with the parameters and coordinates appearing there related to the current ones as follows: $\k^2=-1, s_2 = 2, g(y_2) = \left({l^2 \over 8}\right)^2 - y_2^2;  y_2 = {l^2 \over 8} \cos \theta, \theta^2 = \f$.}. First of all, the 4D base is K\"ahler, with K\"ahler form
\be
\F^3 = -{l^2 \over 8} \left[ \cos \theta \t_2 e^{-2 \F} {i\over 2} dw\wedge d\bar w + \sin\theta d\theta \wedge (d\f +  2 \Im m (\pa \F) )\right]
\ee
It's straightforward to check that $\F^3$ is closed thanks to the equation (\ref{Phieqw}). Adapted complex coordinates can be chosen to be $w$ and the combination
\be W = \ln \sin \theta - \F + i\f .\ee
The  K\"ahler potential is then given by
\be
\calk = \pm {l^2 \over 8}\left[ \sqrt{ 1- e^{2( \Re e (W ) + \F)}} - {\rm arctanh} \sqrt{ 1- e^{2( \Re e (W ) + \F)}}\right]
\ee
where the upper (lower) sign holds in the patch where $\cos \theta>0$ ($\cos \theta <0$) respectively.  Note that, for our rotationally invariant ansatz (\ref{taurotinv}-\ref{Arotinv}), where $\F$ depends
only on $x=\Re e w$, the 4D base is actually a toric K\"ahler manifold, with  $\pa /\pa_\psi$ and $\pa /\pa_\f$ generating the torus action.

In addition to the K\"ahler form $\F^3$, there exist on the base two further selfdual two-forms $\F^1$ and $\F^2$ such that $\F^i, i = 1,2,3$ satisfy the quaternionic algebra. The
 forms $\F^{1,2}$ are covariantly closed with respect to a $U(1)$ connection built
out of the axion-dilaton field. Defining $\F^\pm = \F^1 \pm i \F^2$, they satisfy
\be
d \F^\pm \mp i {d \t_1 \over 2 \t_2} \wedge \F^\pm =0.
\ee
Explicitly, the $\F^\pm$ are given by
\be
\F^+ = {l^2 \over 32}\sqrt{\t_2} e^W dw\wedge d W, \qquad \F^- =\overline{\F^+}  .
\ee
We note that in these solutions the only Killing vector of the 4D metric which also leaves the axion-dilaton profile invariant is $\pa /\pa_\f$. The form $\F^+$ ($\F^-$) is not invariant under the corresponding isometry but carries charge 1 (resp. -1) and the  Killing vector is therefore often called rotational\footnote{The generalization of the analysis of \cite{Bellorin:2006yr} to include solutions with a rotational Killing vector was performed in \cite{Raeymaekers:2014bqa}.}.

The function $f$ and 1-form $\x$, which determine how the time coordinate is fibered, and the vector multiplet fields $Y^I, F^I$  obey a coupled  set of BPS equations\footnote{More precisely, eq. (2.8)-(2.11) in \cite{Raeymaekers:2014bqa}} whose general solution was discussed in \cite{Bena:2004de,Gauntlett:2004qy}.

Now let's turn to the full set of Killing spinors preserved by our solutions describing backreacted M2-particles. We derive these explicitly in Appendix \ref{appKS} to which refer
for more details. It turns out that all our solutions preserve a set of Killing spinors which are constant on the 3D base and depend only on the two-sphere coordinates:
\be
G_{0}^{\b\g} = e^{-{i  \over 2}  \b \f \s_3} e^{ {i \theta \over 2} \g^{\hat \f }} g_{0}^{\b\g}.\label{KSzeromodes}
\ee
Here, $\b, \g = \pm 1$ and the $g_{0}^{\b\g}$ are constant spinors defined in (\ref{spinbasis}).
These Killing spinors are periodic when going around on the boundary cylinder, $\psi \to \psi + 2 \p$, signalling that  our solutions live in the Ramond sector of the dual theory, as does the D4-D0 black hole.
When the M2-charge is turned off, ${q_\star}=0$, our solutions preserve all four $G_{0}^{\b\g}$, which we interpret to correspond to the  zero modes of the four leftmoving supercurrents of the $(4,0)$ theory in the Ramond sector. When the M2-charge is turned on, ${q_\star} \neq 0$, there is an extra projection condition
\be (1 -2  \g_{\hat w \hat \bar w} \s_3 ) G_{0}^{\b\g} =0, \label{M2proj} \ee
which projects on the two Killing spinors (\ref{KSzeromodes}) with $\b = 1$. Note that this is also the projection condition of $\k$-symmetry for a probe M2-particle placed in the $AdS_3$ background \cite{Levi:2009az}.

We propose that the holographic interpretation of the reduction in supersymmetries (\ref{M2proj})  when ${q_\star} \neq 0$ is that the deformation (\ref{MSWdef}) of the boundary CFT breaks the number of preserved left-moving supersymmetries from four to two, and that our solutions represent left-moving Ramond states, preserving the two zero modes of the N=2 supercurrents.
We should also mention that, when the parameter $\m$ in (\ref{taurotinv}-\ref{Arotinv}) is one, there are extra Killing spinors which do depend on  $x_+$: four (resp. two) when ${q_\star}=0$  (${q_\star}\neq 0$). We interpret these as
the extra mode number $\pm 1$  modes of the supercurrents preserved by the Ramond ground state with maximal R-charge, which can be obtained by spectral flow from the NS ground state also preserving eight (resp. four) supercharges.

\section{Outlook}\label{secout}

In this work we reported on progress towards constructing the fully backreacted microstate solutions arising in the black hole deconstruction
proposal. We constructed in detail the  M2-brane solution in the center of AdS and made a concrete proposal for the
 solutions describing M2-branes on helical curves. One of our main intentions was to show that these solutions  are regular away from the M2-brane source, free of closed timelike curves and
asymptotically AdS.

We also studied their holographic interpretation, identifying the dual field theory as a deformation of the MSW theory
and computing the operator VEVs in the states dual to our solutions.
These computations suggest that the solutions  are to be interpreted as Ramond ground states in a dual field theory with a left-moving N=2 superconformal symmetry.  It would be interesting to get a more explicit picture  of said  deformations and states  in  the MSW sigma model \cite{Minasian:1999qn}.

To obtain the 4D ellipsoidal D2-brane solution depicted in Figure \ref{fancyfig}(b) one would like to perform dimensional reduction along $\psi$ on our solutions. This would require some smearing of the M2 charge. One would expect that adding a further probe M2-brane to  our backreacted solution at constant $w$ doesn't break any further supersymmetries and that   it should be possible to smear our solutions on a helical curve  along the $\psi$ direction to obtain a solution which is $\psi$-rotationally invariant. It would be interesting to work this out in detail.

As mentioned in the Introduction the configurations considered here do not include  the backreaction effects of a fundamental string
running between the D6 and anti-D6 centers, which is required by tadpole cancellation. It would be of great interest to add this extra ingredient to our solutions.

Recently in the work \cite{Martinec:2014gka,Martinec:2015pfa} it has been proposed that more generally configurations of M2-branes wrapping the nontrivial cycles in bubbling geometries \cite{Bena:2005va} might possibly be the realization in the gravitational regime of the the pure Higgs states that can be identified in the associated quiver gauge theories \cite{Bena:2012hf, Lee:2012sc}. As our solutions describe the backreaction of a wrapped M2-brane in the simplest bubbling solution (in a decoupling limit), they can also be seen as a first sample calculation in this program.

\section*{Acknowledgements}
We would like to  thank I. Bena, F. Denef, E. Martinec, T. Prochazka and B. Vercnocke for useful comments and discussions.  The research of JR was supported by the Grant Agency of the Czech Republic under the grant
14-31689S. DVdB is partially supported by TUBITAK grant 113F164 and by the Bo\u{g}azi\c{c}i University Research Fund under grant number 13B03SUP7.
This collaboration was supported by the bilateral collaboration grant T\"UBITAK 14/003 \& 114F218.

\bigskip

\appendix

\section{Near-brane expansion}\label{appnb}
In this Appendix we will discuss how to   set up a recursive expansion for the solutions to (\ref{Phieq}) near $x\to  -\infty$ where the M2-particle is located. One finds that  (\ref{Phieq})
is compatible with an $x \to -\infty$  expansion of the form
\be
{e^{\tilde \F}} = e^{-m x}\sum_{n =0}^\infty Q_n (- x^{-1} ) e^{mnx}\label{nearbrane}
\ee
where the positive number $m$ is a first integration constant. The $Q_n (v)$ turn out to be polynomials which can be recursively determined. Upon setting to zero an integration constant which can be absorbed in $m$,
the equation for $Q_{0}$ is solved by
\be
 Q_0 = D
 \ee
 where $D$ is a second integration constant.

  The remaining $Q_n (v) $ are determined recursively by
\begin{align}
& D (k-1) k m^2 v Q_k-D v^3 (-2 k m+m-2 v) \dot Q_k +D v^5 \ddot Q_k \nonu
&=- (v+\epsilon ) \delta _{2,k}-
 \sum _{n=0}^{k-1} \left(m^2 (n-1) v (2 n-k) Q_n Q_{k-n}-m (n-1) v^3 Q_n
  \dot Q_{k-n}\right.\nonu
 & \left. -v^3 (m (k-3 n+1)-2 v) \dot Q_n Q_{k-n} - v^5
   \dot Q_n \dot Q_{k-n} +v^5 \ddot Q_n Q_{k-n}\right)\label{nearrecurs}
\end{align}
   where the dot stands for differentiation with respect to $v$.
   By examining the solutions of the homogeneous equations one sees that their integration constants  can be absorbed in redefinitions of $m$ and $D$.
  One finds that $Q_{2 n+ 1} =0$ and the recursions (\ref{nearrecurs})  for the even $Q_{2n}$ can be easily solved to high order.
   For example, the first  terms are
   \be
  e^{ \tilde \F}=D e^{-m x}+ \frac{x e^{m x} \left(m \left(\epsilon -\frac{1}{x}\right)-\frac{\epsilon
   }{x}\right)}{4 D m^3}+ \calo( e^{3 m x} ).
   \ee

    The asymptotic  condition (\ref{xinfty}) determines the integration constant $m$:
    \be
    m=1.
    \ee
    Comparing with the perturbative solution in $\e$ gives the other integration constant $D$ to order $\e^2$ as:
\be
D = \frac{1}{2}+\frac{\epsilon }{4}+\left(\frac{1}{8}-\frac{\pi ^2}{48}\right) \epsilon ^2 + \calo (\e^3 ).
\ee

\section{Holographic renormalization for 3D axion-dilaton gravity}\label{apphol}
Here we discuss the holographic renormalization  for the 3D axion-dilaton theory defined by (\ref{3daction}). The analysis is largely similar to the one for a
 massless scalar coupled to 3D gravity which was discussed in \cite{de Haro:2000xn}, while a discussion of holographic renormalization for higher-dimensional axion-dilaton
 theories appears in \cite{Papadimitriou:2011qb},\cite{Chemissany:2012du}.
Setting $\t_2 = e^{-\Psi}$, we start from  the  3D action:
\bea
S &=& {1 \over 16 \p G_3} \int_\calm \left[  d^3 x \sqrt{- G} \left( \calr + {2 \over l^2} -\half \pa_\a \Psi \pa^\a \Psi- {e^{2 \Psi}\over 2}\pa_\a \t_1 \pa^\a \t_1 + {l \over 2}  \cala\wedge d \cala \right)\right.\nonu
&&\left. - 2 \int_{\d \calm} \sqrt{-\g} K \right]\label{Sbare}
\eea
leading to the equations of motion
\bea
\cale_{\a\b} &=& \calr_{\a\b} + {2 \over l^2} G_{\a\b} - \half \pa_\a \Psi \pa_\b \Psi- {e^{2 \Psi}\over 2}\pa_\a \t_1 \pa_\b \t_1=0\\
\cale_{\Psi} &=& \Box \Psi - e^{2 \Psi} (\pa\t_1)^2 =0\\
\cale_{\t_1} &=& \nabla^\a \left( e^{2 \Psi}\pa_\a \t_1 \right) =0\label{eom3D}
\eea
We use a coordinate system in terms of which the metric looks like
\be
ds^2_3 = l^2 \left( {dy^2 \over 4 y^2} + {1 \over y} g_{ij} (x^k,y) dx^i dx^j\right)
\ee
 and assume the standard
Fefferman-Graham expansion \cite{FG} for the fields near the boundary:
\bea
g_{ij} &=&g_{(0)ij}+ yg_{(2)ij}+ y\ln y\, \tilde g_{(2)ij} + \calo ( y^2 \ln y)\label{FG1}\\
\Psi &=& \Psi_{(0)} + y\Psi_{(2)} + y\ln y\, \tilde \Psi_{(2)}+ \calo ( y^2 \ln y)\\
\t_1 &=& \t_{1(0)} + y\t_{1(2)} + y\ln y\, \tilde \t_{1(2)}+ \calo ( y^2 \ln y)\\
\cala &=& \cala_{(0)} + \calo (y).\label{FG4}
\eea
Substituting these in the equations  of motion (\ref{eom3D}) and working out the leading terms one finds that the logarithmic coefficients
$\tilde g_{(2)}, \tilde \Psi_{(2)}, \tilde \t_{1(2)}$ are completely determined  by the boundary values $ g_{(0)},  \Psi_{(0)},  \t_{1(0)}$:
\bea
\tilde g_{(2)ij} &=& - {1\over 4} \left( \pa_i \Psi_{(0)} \pa_j \Psi_{(0)}  + {e^{2 \Psi_{(0)}}} \pa_i \t_{1(0)} \pa_j \t_{1(0)}\right)
+  {1\over 8} \left( (\pa \Psi_{(0)})^2  + {e^{2 \Psi_{(0)}}} (\pa \t_{1(0)})^2 \right) g_{(0)ij} \nonu
\tilde \Psi_{(2)} &=&  - {1\over 4} \Box  \Psi_{(0)}  + {e^{2 \Psi_{(0)}} \over 4} (\pa \t_{1(0)})^2 \\
\tilde \t_{1(2)} &=&  - {1\over 4} \Box  \t_{1(0)}   - \half \pa_i \Psi_{(0)} \pa^i \t_{1(0)}\label{tildesols}
\eea
where indices are raised and covariant derivatives taken with respect to the boundary metric $g_{(0)}$. We note that $\tilde g_{(2)ij}$ is traceless.

For the tensor
 $g_{(2)ij}$  on the other hand, only the trace and divergence are fixed by $ g_{(0)},  \Psi_{(0)},  \t_{1(0)}$:
 \bea
 g_{(2)} &=& - \half R_{(0)} +  {1\over 4} \left( (\pa \Psi_{(0)})^2  + {e^{2 \Psi_{(0)}}} (\pa \t_{1(0)})^2 \right)\\
 \nabla^j  g_{(2)ij}&=& \pa_i  g_{(2)} + \Psi_{(2)}\pa_i \Psi_{(0)}+  \t_{1(2)} \pa_i  \t_{1(0)}\label{g2sol}
 \eea
 As we shall see below, $g_{(2)ij}$ essentially encodes  the expectation value of
  the CFT stress tensor. The functions    $\Psi_{(2)},  \t_{1(2)}$ are completely free and encode the expectation values of
  the operators dual to $\Psi, \t_1$ respectively. As usual, these undetermined modes are fixed by physical requirements on the solution in the  interior, such as regularity or, in our
  case, matching onto the proper source term.

  Proceeding as in \cite{de Haro:2000xn}, we regularize the action by cutting off the $y$ integral at $y= \d\ll 1$.
  One finds for the regularized on-shell action
  \be
  S_{\mathrm{reg}} = - {l \over 8 \p G_3} \int d^2 x \left[  \int_\d dy{\sqrt{-g }\over y^2}  + 2\left.\left( \pa_y\sqrt{-g } - {\sqrt{-g }\over y}\right)\right|_{y=\d} \right].\label{Sreg}
  \ee
  Using (\ref{FG1}) and (\ref{g2sol}) one finds that this contains the following divergent terms as $\d \to 0 $:
  \be
  S_{\mathrm{div}} = {l \over 16 \p G_3}\int_{\d \calm}  d^2x\sqrt{- g_{(0)}}\left[  {2 \over \d }- \half \left( R_{(0)} - \half (\pa \Psi_{(0)})^2  - \half {e^{2 \Psi_{(0)}}} (\pa \t_{1(0)})^2\right)\ln \d \right]
  \ee
  These divergences are cancelled if we add the
  boundary counterterm action
  \be
  S_{\mathrm{ct}} = {l \over 16 \p G_3}\int_{\d \calm}  d^2x\sqrt{- g}\left[ - {2 \over \d }+ \half \left( R - \half (\pa \Psi)^2  - \half {e^{2 \Psi}} (\pa \t_{1})^2\right)\ln \d \right]
  \ee
  We could of course have added further local finite terms as $\d \to 0$, which in the dual field theory corresponds to using a  different renormalization scheme.
   We then obtain the renormalized action
  \be
  S_{\mathrm{ren}} = S_{\mathrm{reg}} + S_{\mathrm{ct}} .
  \ee
We can now determine the holographic one-point functions of various operators in our solution by varying the renormalized action with respect to the boundary sources.
Here we must be careful to vary the orginal action (\ref{Sbare}) and not the expression (\ref{Sreg}), which differs from it by terms proportional to the equations of motion whose variation is however not zero.

 By varying with respect to the boundary metric we obtain the contribution to the expectation value of the stress tensor  coming from bulk fields coupling to the metric (as we will
 see below, there is an extra contribution from the Chern-Simons gauge field):
 \be
 \langle T_{ij}^{\mathrm{grav}} \rangle= - \lim_{\d \to 0}{2 \over \sqrt{- g} } {\d S_{\mathrm{ren}} \over \d  g^{ij}} = \lim_{\d \to 0}\left( T_{ij}^{\mathrm{reg}} + T_{ij}^{\mathrm{ct}} \right)
 \ee
 We find
 \bea
 T_{ij}^{\mathrm{reg}} &=& {1 \over 8 \p G_3} (K_{ij} - K \g_{ij})\\
 &=& {l \over 8 \p G_3}\left( g_{ij}' + {g_{ij} \over y}- g^{kl} g_{kl}' g_{ij}\right)_{|y= \d}\\
 &=&  {l \over 8 \p G_3}\left(  {g_{(0)ij} \over \d} + 2   \tilde g_{(2)ij}\ln \d +  2 g_{(2)ij}+ \tilde g_{(2)ij}- g_{(2)}g_{(0)ij}\right)\\
 T_{ij}^{\mathrm{ct}} &=&  {l \over 8 \p G_3}\left( - {g_{(0)ij} \over \d} - 2   \tilde g_{(2)ij}\ln \d - g_{(2)ij}\right)
 \eea
 so that we obtain for the renormalized stresstensor
 \be
\langle  T_{ij}^{\mathrm{grav}} \rangle = {l \over 8 \p G_3}\left( g_{(2)ij}+ \tilde g_{(2)ij}- g_{(2)}g_{(0)ij}\right)\label{stresst}
  \ee

 As reviewed in detail in \cite{Kraus:2006wn}, the inclusion of the Chern-Simons field $\cala$ gives an extra contribution
 to the stress tensor. Since the two components of $\cala$ are conjugate variables, we are to hold fixed only one of them, say $\cala_-$, on the boundary.
  A correct variational principle for this boundary condition requires the addition of a metric-dependent boundary term
  \be
  S^{\mathrm{ct}}_\cala = - {l \over 64 \p G_3} \int_{\d \calm} d^2x\sqrt{-g} g^{ij} \cala_i \cala_j\label{Act}
  \ee
 which gives a contribution to the boundary stress tensor
 \be
 \langle T_{ij}^{\cala} \rangle= {l \over 32 \p G_3} \left( \cala_{(0)i}\cala_{(0)j}- \half \cala_{(0)}^k \cala_{(0)k} g_{(0)ij} \right).
 \ee
 The total stress tensor is then $ \langle T_{ij}\rangle= \langle T_{ij}^{\mathrm{grav}} \rangle +  \langle T_{ij}^{\cala} \rangle$. We note that the trace anomaly
 $ \langle T_{i}^i\rangle$ is proportional to $g_{(2)}$ given in (\ref{tranom}):
 \be
  \langle T_{i}^i\rangle = -{l \over 32 \p G_3}\left( (\pa \Psi_{(0)})^2  + {e^{2 \Psi_{(0)}}} (\pa \t_{1(0)})^2 - 2 R_{(0)} \right)\label{tranom}
  \ee

The operator dual to $\cala$ is an R-symmetry current $J^3$, and one finds for its holographic one point function
 \be
 \langle J^3_i  \rangle\equiv -\lim_{\d \to 0}{1 \over \sqrt{- g} } {\d S_{\mathrm{ren}} \over \d \cala^i}  =  {l \over 16 \p G_3}\cala_{(0)i}.
 \ee

 The variation of the renormalized action with respect to $\Psi$ and $\t_1$ gives the renormalized one-point functions of the dual operators $\calo_\Psi$ and $\calo_{\t_1}$:
 \bea
 \langle\calo_\Psi \rangle^{\mathrm{ren}} &=& - \lim_{\d \to 0}{1 \over \sqrt{- g} } {\d S_{\mathrm{ren}} \over \d \Psi}= \lim_{\d \to 0}\left(  \langle\calo_\Psi \rangle^{\mathrm{reg}}+\langle\calo_\Psi \rangle^{\mathrm{ct}} \right)\\
 \langle\calo_{\t_1} \rangle^{\mathrm{ren}} &=& - \lim_{\d \to 0}{1 \over \sqrt{- g} } {\d S_{\mathrm{ren}} \over \d \t_1}= \lim_{\d \to 0}\left(  \langle\calo_{\t_1} \rangle^{\mathrm{reg}}+\langle\calo_{\t_1} \rangle^{\mathrm{ct}} \right)
 \eea
 One finds
 \begin{align}
  \langle\calo_\Psi \rangle^{\mathrm{reg}} =&- {l \over 8 \p  G_3}\left( \Psi_{(2)} + ( 1+ \ln \d) \tilde \Psi_{(2)}\right), &
  \langle\calo_\Psi \rangle^{\mathrm{ct}} =&{l \over 8 \p  G_3} \ln \d \tilde \Psi_{(2)} \\
  \langle\calo_{\t_1} \rangle^{\mathrm{reg}} =& - {l e^{2 \Psi_{(0)}} \over 8 \p  G_3}\left( \t_{1 (2)} + ( 1+ \ln \d) \tilde \t_{1(2)}\right),&
  \langle\calo_{\t_1} \rangle^{\mathrm{ct}} =&    {l e^{2 \Psi_{(0)}} \over 8 \p  G_3} \ln \d \tilde \t_{1(2)}
 \end{align}
 with the upshot
  \bea
  \langle\calo_\Psi \rangle^{\mathrm{ren}} &=&- {l \over 8 \p  G_3}\left( \Psi_{(2)} + \tilde \Psi_{(2)}\right)\\
  \langle\calo_{\t_1} \rangle^{\mathrm{ren}} &=& - {l e^{2 \Psi_{(0)}} \over 8 \p  G_3}\left( \t_{1 (2)} + \tilde \t_{1(2)}\right)
  \eea

One further result we will need in the main text is the following. Taking the covariant derivative of the boundary stress tensor, we find the following relation
\be
\nabla^i \langle T_{ij} \rangle = -  \langle\calo_\Psi \rangle \pa_j \Psi_{(0)} - \langle\calo_{\t_1} \rangle \pa_j \t_{1 (0)}+ \nabla^i \langle T_{ij}^{\cala} \rangle \label{Wardid}
\ee
which reflects a  Ward identity for the breaking of conformal invariance in the presence of sources in the dual theory \cite{de Haro:2000xn}.
\section{Review of 5D axion-dilaton solutions}\label{app5D}
In this Appendix we review the 5D action and Killing spinor equations relevant for our solutions.
We consider 11-dimensional supergravity compactified on a Calabi-Yau threefold with triple intersection form $D_{IJK}$. Dimensionally reducing to 5D gives  ungauged  N=1 supergravity coupled to $h_{(1,1)}-1$ vector multiplets and
 $h_{(2,1)}+1$ hypermultiplets, the action for which can be found in\footnote{To obtain  our conventions from those used in ref. \cite{Bellorin:2006yr}, one should send $g_{\m\n} \to - g_{\m\n},\ \g_\m \to i \g_\m, \ \g^\m \to -i \g^\m$ to account for the fact that our metric signature   is mostly plus, and replace the quantities $h^I$ and $C_{IJK}$ in \cite{Bellorin:2006yr} by  $h^I \to Y^I/\sqrt{3},\ C_{IJK} \to \sqrt{3}D_{IJK}/2$. We denote tangent space indices with a hat, and use a representation   where $\g^{\hat 0 , \hat 1,\hat 2,\hat 3}$ are real and $\g^{\hat  4}$ is imaginary, satisfying $\g^{\hat 0 \hat 1\hat 2\hat 3 \hat 4}=i$.}
  \cite{Bellorin:2006yr}.
 We make a consistent truncation
of this theory where the vector multiplet scalars $Y^I$ (normalized such that $D_{IJK} Y^I Y^J Y^K =6$) are constant, and where all hypermultiplet scalars,
apart from the axion-dilaton field $\t$ in the universal hypermultiplet, are constant as well. The truncated theory has bosonic action
\bea
S &=& \int d^5 x \sqrt{-g} \left[ R - {\pa_\m \t \pa^\m \bar \t \over 2  \t_2^2} -  {1 \over 4}a_{IJ} F^I_{\m\n} F^{J \m\n}\right]+ {DIJK  \over 6} \int F^I\wedge F^J \wedge A^K\\
a_{IJ} &=& - D_{IJK} Y^K + {1\over 4} Y_I Y_J, \qquad Y_I =  D_{IJK} Y^J Y^K \label{5daction}
\eea
The resulting equations of motion are
\bea
R_{\m\n} -  {\pa_{(\m} \t \pa_{\n )} \bar \t \over 2 \t_2^2} +{1\over 12} g_{\m\n} a_{IJ} F^I_{\r\s} F^{J \r\s}-{1 \over 2} a_{IJ} F^I_{\m\r} F^{J \ \r}_{\ \n} &=&0\\
\partial_\m\left(\sqrt{-g}g^{\m\n}\partial_\n\tau\right) + i\sqrt{-g}g^{\m\m}\frac{\partial_\m\tau\partial_\m\tau}{\tau_2}&=&0\\
a_{IJ} d (\star F^J ) + {D_{IJK}\over 2} F^J \wedge F^K &=&0\label{eom}
\eea
Now let's discuss the supersymmetry variations of the fields. The supersymmetry parameter consists of two complex  4-component spinors
\be \e = \left( \begin{array}{c} \e^1 \\ \e^2 \end{array} \right),\ee  which are related by a symplectic Majorana condition $\G_\calm \cdot \e = \e$, where $\G_\calm $ is an idempotent operator acting as
\be
\G_\calm \cdot \e = \g^{\hat{4}} \s_2 \e^\star \label{symplMaj}
\ee
and $\star$ denotes complex conjugation.
After imposing the symplectic Majorana condition, $\e$ contains
 a total of 8 independent real components.

The Killing spinor equations, i.e. the vanishing of the supersymmetry variations of the gravitino, gauginos and hyperinos, reduce to, respectively:
\bea
\left[ \nabla_\m + {i Y_I \over 48}\left( F^{I\n\r}\g_{\m\n\r} F^I - 4 F^{I\ \n}_{\ \m} \g_\n \right) + i { \pa_\m \t_1 \over 4 \t_2} \s_3 \right]\e &=& 0\label{gravitino}\\
a_{IJ} \slashed F^I{\pa Y^J\over \pa \f^x} \e &=&0\label{gaugino}\\
\slashed \pa \bar \t \e^1 = \slashed \pa  \t \e^2 &=&0\label{hyperino}
\eea
where
$\f^x$ are $n_V$ coordinates on the surface $D_{IJK} Y^I Y^J Y^K =6$.

\section{Killing spinors}\label{appKS}
In this Appendix we discuss  how many supersymmetries are preserved by our solutions (\ref{taurotinv}-\ref{Arotinv})  for general values of the parameter $\m$,
 and give explicit expressions for the Killing spinors. We look for supersymmetry parameters $G$ for which the
 gravitino, gaugino and hyperino variations (\ref{gravitino}-\ref{hyperino}) vanish. It's easy to see that the gaugino variation is automatically zero since the gauge fields are of the form $F^I \sim Y^I F$.
  Choosing the vielbein
  \begin{align}
  e^{\hat t} =&{l \over 2}(d (t+ \m \psi)  + \F'  d\psi ) &  e^{\hat \theta } =& {l \over 2} d \theta\\
   e^{\hat x} =& {l \over 2} \sqrt{\t_2} e^{ - \F} dx &  e^{\hat \f } =& {l \over 2} d (\f - t -  \m \psi)\\
    e^{\hat \psi} =& {l \over 2} \sqrt{\t_2} e^{ - \F} d\psi & &
    \end{align}
one finds, using  (\ref{Phieq}), the following nonvanishing spin connection components:
\begin{align}
  \o^{\hat t \hat x} =& - \half \sqrt{\t_2} e^{ - \F} d\psi , &
   \o^{\hat t \hat \psi} =&  \half \sqrt{\t_2} e^{ - \F} dx \\
    \o^{\hat x \hat \psi} =& - \half d (t+\m \psi)   + \half \left( \F' + {{q_\star} \over \t_2} \right) d\psi &\o^{\hat \theta \hat \f} =& - \cos \theta d (\f - t -  \m \psi)
   \end{align}

After some algebra the  gravitino and hyperino equations can then be rewritten as
\bea
\left[ U \pa_\m U^{-1} + {q_\star \d_\m^\psi \g^{\hat x \hat \psi} \over 4 \t_2} \left(1 - i \g^{\hat x \hat \psi} \s_3 \right) \right] G &=& 0\\
{q_\star} \left(1 -  \g^{\hat x \hat \psi} \s_3 \right) G &=& 0\label{M2projapp}
\eea
where
\be
U = e^{ {i \theta \over 2} \g^{\hat \f }}e^{{\f- t -  \m \psi \over 2}\g^{\hat \theta \hat \f} }e^{{ t+\m \psi  \over 2}\g^{\hat x \hat \psi} }
\ee
From this we conclude that the solutions (\ref{taurotinv}-\ref{Arotinv}) preserve local Killing spinors  of the form \be G = U \cdot g, \label{KS1} \ee where $g$ is a constant spinor satisfying the symplectic
Majorana condition (\ref{symplMaj}). When the M2-charge ${q_\star}$ is nonzero, $g$ should in addition satisfy the projection condition (\ref{M2projapp}).

It will be useful to choose a specific basis of constant spinors by diagonalizing 3 commuting idempotent operators which also commute with the operator $\G_\calm$ appearing in the symplectic Majorana condition (\ref{symplMaj}).
We label these basis elements as $g_{1-\a\over 2}^{\b\g}$, with $\a ,\b ,\g = \pm 1 $ and take them to satisfy
\bea
i \g^{\hat t} g_{1-\a\over 2}^{\b\g} &=& \a g_{1-\a\over 2}^{\b\g}\\
i \g^{\hat \theta \hat \f} \s_3 g_{1-\a\over 2}^{\b\g} &=& \b g_{1-\a\over 2}^{\b\g}\\
i \g^{\hat \psi \hat \f } \s_1 g_{1-\a\over 2}^{\b\g} &=& \g g_{1-\a\over 2}^{\b\g}.\label{spinbasis}
\eea
Spinors of the form (\ref{KS1}) can then be rewritten as
\be
G_{1-\a\over 2}^{\b\g} = e^{{i  \over 2} \left[ \b (1-\a) (t + \m \psi ) - \b \f \right]\s_3} e^{ {i \theta \over 2} \g^{\hat \f }} g_{1-\a\over 2}^{\b\g}.\label{KS2}
\ee

For $0<\m <1$, only the  local Killing spinors which obey $\a=1$ have a  well-defined global periodicity under $\psi \to \psi + 2\p$. In particular they are periodic on the
boundary cylinder and should be interpreted as belonging to the Ramond sector of boundary theory. Explicitly they are given by
\be
G_{0}^{\b\g} = e^{-{i  \over 2}  \b \f \s_3} e^{ {i \theta \over 2} \g^{\hat \f }} g_{0}^{\b\g}.\label{zeromodeKS}
\ee
In the absence of an M2-particle, these are to be interpreted as the 4 zero modes of the supercurrents in the $(4,0)$ algebra  preserved by  the Ramond ground states. When the
M2-charge ${q_\star}$ is nonzero, we have to impose the extra projection condition (\ref{M2proj}) which sets $\b =1$ and the solution has only two Killing spinors $G_0^{+ \g}$.

The case $\m=1$ is special, as then even the spinors $G_{1-\a\over 2}^{\b\g}$ with $\a = -1$ are periodic under $\psi \to \psi + 2\p$. Without M2-charge, these four extra Killing spinors reflect the fact that the Ramond ground state with maximal R-charge preserves four nonzero modes of the supercurrents (as can be easily seen from its interpretation as the spectral flow of the NS ground state which is maximally supersymmetric). When ${q_\star}\neq 0$, the projection (\ref{M2proj}) condition imposes $\a = \b$, so that we then preserve the four Killing spinors\footnote{In terms of the light-cone coordinates on the boundary, the Killing spinors
$G_1^{- \g}$ are given by $$ G_1^{- \g}= e^{-{i \s_2 } \left[ (1- {\e\over 2} ) x_+ +  {\e\over 2} x_- + {\b \f \over 2} \right]} e^{ {i \theta \over 2} \g^{\hat \f }} g_{1}^{-\g}.$$
The  non-integer mode numbers in this expression are puzzling and we have not understood them at  present.}  $G_0^{+ \g}, G_1^{- \g}$.

This analysis is now easily extended to the more general solutions describing an M2-particle on a helical curve constructed in section \ref{secM2spiral}.
Since these are obtained by performing a large coordinate transformation on the 3D coordinates $(y,x_+,x_-)$, these solutions preserve the same zero mode Killing spinors
(\ref{zeromodeKS}). Once again, for $\m=1$ there are additional Killing spinors which do depend on the 3D coordinates.


\begin{thebibliography}{99}

\bibitem{Strominger:1996sh}
  A.~Strominger and C.~Vafa,
  ``Microscopic origin of the Bekenstein-Hawking entropy,''
  Phys.\ Lett.\ B {\bf 379}, 99 (1996)
  [hep-th/9601029].

  \bibitem{Mathur:2005zp}
  S.~D.~Mathur,
  ``The Fuzzball proposal for black holes: An Elementary review,''
  Fortsch.\ Phys.\  {\bf 53}, 793 (2005)
  [hep-th/0502050].

   \bibitem{Bena:2013dka}
  I.~Bena and N.~P.~Warner,
  ``Resolving the Structure of Black Holes: Philosophizing with a Hammer,''
  arXiv:1311.4538 [hep-th].

 \bibitem{Denef:2007yt}
  F.~Denef, D.~Gaiotto, A.~Strominger, D.~Van den Bleeken and X.~Yin,
  ``Black Hole Deconstruction,''
  JHEP {\bf 1203}, 071 (2012)
  [hep-th/0703252 [HEP-TH]].



\bibitem{Maldacena:1997de}
  J.~M.~Maldacena, A.~Strominger and E.~Witten,
  ``Black hole entropy in M theory,''
  JHEP {\bf 9712}, 002 (1997)
  [hep-th/9711053].

\bibitem{Levi:2009az}
  T.~S.~Levi, J.~Raeymaekers, D.~Van den Bleeken, W.~Van Herck and B.~Vercnocke,
  ``Godel space from wrapped M2-branes,''
  JHEP {\bf 1001}, 082 (2010)
  [arXiv:0909.4081 [hep-th]].

 \bibitem{deBoer:2008fk}
  J.~de Boer, F.~Denef, S.~El-Showk, I.~Messamah and D.~Van den Bleeken,
  ``Black hole bound states in AdS(3) x S**2,''
  JHEP {\bf 0811}, 050 (2008)
  [arXiv:0802.2257 [hep-th]].

 \bibitem{Minasian:1999qn}
  R.~Minasian, G.~W.~Moore and D.~Tsimpis,
  ``Calabi-Yau black holes and (0,4) sigma models,''
  Commun.\ Math.\ Phys.\  {\bf 209}, 325 (2000)
  [hep-th/9904217].










\bibitem{Bates:2003vx}
  B.~Bates and F.~Denef,
  ``Exact solutions for supersymmetric stationary black hole composites,''
  JHEP {\bf 1111}, 127 (2011)
  [hep-th/0304094].

\bibitem{deBoer:2008zn}
  J.~de Boer, S.~El-Showk, I.~Messamah and D.~Van den Bleeken,
  ``Quantizing N=2 Multicenter Solutions,''
  JHEP {\bf 0905}, 002 (2009)
  [arXiv:0807.4556 [hep-th]].

\bibitem{Myers:1999ps}
  R.~C.~Myers,
  ``Dielectric branes,''
  JHEP {\bf 9912}, 022 (1999)
  [hep-th/9910053].

  \bibitem{Gaiotto:2004ij}
  D.~Gaiotto, A.~Strominger and X.~Yin,
  ``Superconformal black hole quantum mechanics,''
  JHEP {\bf 0511}, 017 (2005)
  [hep-th/0412322].

 \bibitem{Brodie:2000yz}
  J.~H.~Brodie, L.~Susskind and N.~Toumbas,
  ``How Bob Laughlin tamed the giant graviton from Taub - NUT space,''
  JHEP {\bf 0102}, 003 (2001)
  [hep-th/0010105].


\bibitem{Raeymaekers:2014bqa}
  J.~Raeymaekers and D.~Van den Bleeken,
  ``Unlocking the Axion-Dilaton in 5D Supergravity,''
  arXiv:1407.5330 [hep-th].

\bibitem{Greene:1989ya}
  B.~R.~Greene, A.~D.~Shapere, C.~Vafa and S.~T.~Yau,
  ``Stringy Cosmic Strings and Noncompact Calabi-Yau Manifolds,''
  Nucl.\ Phys.\ B {\bf 337}, 1 (1990).

\bibitem{Maldacena:2000dr}
  J.~M.~Maldacena and L.~Maoz,
  ``Desingularization by rotation,''
  JHEP {\bf 0212}, 055 (2002)
  [hep-th/0012025].

 \bibitem{Simons:2004nm}
  A.~Simons, A.~Strominger, D.~M.~Thompson and X.~Yin,
  ``Supersymmetric branes in AdS(2) x S**2 x CY(3),''
  Phys.\ Rev.\ D {\bf 71}, 066008 (2005)
  [hep-th/0406121].

\bibitem{Gaiotto:2004pc}
  D.~Gaiotto, A.~Simons, A.~Strominger and X.~Yin,
  ``D0-branes in black hole attractors,''
  hep-th/0412179.

 \bibitem{Schwimmer:1986mf}
  A.~Schwimmer and N.~Seiberg,
  ``Comments on the N=2, N=3, N=4 Superconformal Algebras in Two-Dimensions,''
  Phys.\ Lett.\ B {\bf 184}, 191 (1987).

  \bibitem{David:1999zb}
  J.~R.~David, G.~Mandal, S.~Vaidya and S.~R.~Wadia,
  ``Point mass geometries, spectral flow and AdS(3) - CFT(2) correspondence,''
  Nucl.\ Phys.\ B {\bf 564}, 128 (2000)
  [hep-th/9906112].

   \bibitem{Braun:2008ua}
  A.~P.~Braun, A.~Hebecker and H.~Triendl,
  ``D7-Brane Motion from M-Theory Cycles and Obstructions in the Weak Coupling Limit,''
  Nucl.\ Phys.\ B {\bf 800}, 298 (2008)
  [arXiv:0801.2163 [hep-th]].

\bibitem{Bergshoeff:2006jj}
  E.~A.~Bergshoeff, J.~Hartong, T.~Ortin and D.~Roest,
  ``Seven-branes and Supersymmetry,''
  JHEP {\bf 0702}, 003 (2007)
  [hep-th/0612072].

  \bibitem{Michel:2014lva}
    B.~Michel, E.~Mintun, J.~Polchinski, A.~Puhm and P.~Saad,
    ``Remarks on brane and antibrane dynamics,''
    JHEP {\bf 1509}, 021 (2015)
    [arXiv:1412.5702 [hep-th]].

\bibitem{Raeymaekers:2011uz}
  J.~Raeymaekers,
  ``Chronology protection in stationary three-dimensional spacetimes,''
  JHEP {\bf 1111}, 024 (2011)
  [arXiv:1106.5098 [hep-th]].

 \bibitem{Mars:1992cm}
  M.~Mars and J.~M.~M.~Senovilla,
  ``Axial symmetry and conformal Killing vectors,''
  Class.\ Quant.\ Grav.\  {\bf 10}, 1633 (1993)
  [gr-qc/0201045].

\bibitem{FG}
C. Fefferman and C. R. Graham, ``Conformal invariants,'' in The Mathematical Heritage
of Elie Cartan (Lyon, 1984), Ast\'erisque, 1985, Numero Hors Serie, 95–116.

\bibitem{de Haro:2000xn}
  S.~de Haro, S.~N.~Solodukhin and K.~Skenderis,
  ``Holographic reconstruction of space-time and renormalization in the AdS / CFT correspondence,''
  Commun.\ Math.\ Phys.\  {\bf 217}, 595 (2001)
  [hep-th/0002230].

  \bibitem{Costa:2010cn}
  R.~N.~Caldeira Costa and M.~Taylor,
  ``Holography for chiral scale-invariant models,''
  JHEP {\bf 1102}, 082 (2011)
  [arXiv:1010.4800 [hep-th]].

\bibitem{Balasubramanian:2009bg}
  V.~Balasubramanian, J.~de Boer, M.~M.~Sheikh-Jabbari and J.~Simon,
  ``What is a chiral 2d CFT? And what does it have to do with extremal black holes?,''
  JHEP {\bf 1002}, 017 (2010)
  [arXiv:0906.3272 [hep-th]].

\bibitem{Detournay:2012pc}
  S.~Detournay, T.~Hartman and D.~M.~Hofman,
  ``Warped Conformal Field Theory,''
  Phys.\ Rev.\ D {\bf 86}, 124018 (2012)
  [arXiv:1210.0539 [hep-th]].

\bibitem{Cheb-Terrab}
 E. S. Cheb-Terrab and A. D. Roche, ``Abel ODEs: Equivalence and integrable classes''
Computer Physics Communications {\bf 130}, 204 (2000) [math-ph/0001037]



\bibitem{Bellorin:2006yr}
  J.~Bellorin, P.~Meessen and T.~Ortin,
  ``All the supersymmetric solutions of N=1,d=5 ungauged supergravity,''
  JHEP {\bf 0701}, 020 (2007)
  [hep-th/0610196].











 \bibitem{Cadavid:1995bk}
  A.~C.~Cadavid, A.~Ceresole, R.~D'Auria and S.~Ferrara,
  ``Eleven-dimensional supergravity compactified on Calabi-Yau threefolds,''
  Phys.\ Lett.\ B {\bf 357}, 76 (1995)
  [hep-th/9506144].

 \bibitem{Gauntlett:2002nw}
  J.~P.~Gauntlett, J.~B.~Gutowski, C.~M.~Hull, S.~Pakis and H.~S.~Reall,
  ``All supersymmetric solutions of minimal supergravity in five- dimensions,''
  Class.\ Quant.\ Grav.\  {\bf 20}, 4587 (2003)
  [hep-th/0209114].
























\bibitem{Bena:2004de}
  I.~Bena and N.~P.~Warner,
  ``One ring to rule them all ... and in the darkness bind them?,''
  Adv.\ Theor.\ Math.\ Phys.\  {\bf 9}, 667 (2005)
  [hep-th/0408106].

\bibitem{Gauntlett:2004qy}
  J.~P.~Gauntlett and J.~B.~Gutowski,
  ``General concentric black rings,''
  Phys.\ Rev.\ D {\bf 71}, 045002 (2005)
  [hep-th/0408122].

\bibitem{Martinec:2014gka}
  E.~J.~Martinec,
  ``The Cheshire Cap,''
  JHEP {\bf 1503}, 112 (2015)
  [arXiv:1409.6017 [hep-th]].

\bibitem{Martinec:2015pfa}
  E.~J.~Martinec and B.~E.~Niehoff,
  ``Hair-brane Ideas on the Horizon,''
  arXiv:1509.00044 [hep-th].

   \bibitem{Bena:2005va}
    I.~Bena and N.~P.~Warner,
    ``Bubbling supertubes and foaming black holes,''
    Phys.\ Rev.\ D {\bf 74}, 066001 (2006)
    [hep-th/0505166].

    \bibitem{Bena:2012hf}
  I.~Bena, M.~Berkooz, J.~de Boer, S.~El-Showk and D.~Van den Bleeken,
  ``Scaling BPS Solutions and pure-Higgs States,''
  JHEP {\bf 1211}, 171 (2012)
  [arXiv:1205.5023 [hep-th]].

\bibitem{Lee:2012sc}
  S.~J.~Lee, Z.~L.~Wang and P.~Yi,
  ``Quiver Invariants from Intrinsic Higgs States,''
  JHEP {\bf 1207}, 169 (2012)
  [arXiv:1205.6511 [hep-th]].



\bibitem{Papadimitriou:2011qb}
  I.~Papadimitriou,
  ``Holographic Renormalization of general dilaton-axion gravity,''
  JHEP {\bf 1108}, 119 (2011)
  [arXiv:1106.4826 [hep-th]].

\bibitem{Chemissany:2012du}
  W.~Chemissany, D.~Geissbuhler, J.~Hartong and B.~Rollier,
  ``Holographic Renormalization for z=2 Lifshitz Space-Times from AdS,''
  Class.\ Quant.\ Grav.\  {\bf 29}, 235017 (2012)
  [arXiv:1205.5777 [hep-th]].

\bibitem{Kraus:2006wn}
  P.~Kraus,
  ``Lectures on black holes and the AdS(3) / CFT(2) correspondence,''
  Lect.\ Notes Phys.\  {\bf 755}, 193 (2008)
  [hep-th/0609074].














\end{thebibliography}
\end{document}